\documentclass{iopjournal}

\usepackage{amsmath,amssymb,amsfonts}

\usepackage{graphicx}
\usepackage{subcaption}

\usepackage{multirow}
\usepackage{booktabs}
\usepackage{bigstrut}
\usepackage{blkarray}

\usepackage{physics}
\usepackage{microtype}

\usepackage{appendix}
\usepackage{cite}
\usepackage{float}
\usepackage[normalem]{ulem}

\begin{document}

\articletype{PAPER} 

\title{Tracking Entanglement Transfer: Emergence of Thermodynamics from Quantum Information}

\author{Debraj Debata$^{1,*}$\orcid{0009-0009-1356-8889}, 
Abhirup Mukherjee$^1$\orcid{0000-0001-7869-7205} and 
Siddhartha Lal$^1$\orcid{0000-0002-5387-6044}}

\affil{$^1$Department of Physical Sciences, Indian Institute of Science Education and Research Kolkata, Mohanpur Campus, West Bengal- 741246, India}

\affil{$^*$Author to whom any correspondence should be addressed.}

\email{dd20ip014@iiserkol.ac.in}
\email{am18ip014@iiserkol.ac.in}
\email{slal@iiserkol.ac.in}


\begin{abstract}
We study entanglement transfer in a minimal model of two qubits that are coupled through an antiferromagnetic Heisenberg exchange ($\tilde{J}$), and where one of them is additionally coupled to a fermionic environment through another antiferromagnetic Heisenberg exchange ($J_{K}$). By tuning the coupling ratio $J_{K}/\tilde{J}$, the system undergoes a quantum phase transition at \(T=0\), accompanied by a redistribution of entanglement from the two-qubit-subsystem to the environment. Remarkably, the resulting physics exhibits properties that bear clear analogy with a quantum-information theoretic perspective of the physics of black hole thermodynamics and continuous Hawking-Page phase transitions. Carefully selected bipartite and tripartite mutual information measures display behaviour analogous to the dynamical evolution of black hole entropy, Hawking entropy, and the Page curve under renormalisation as well as temporal dynamics that is expected during the process of evaporation. An effective temperature scale is obtained from the variation of the ground state energy with respect to changes in the bipartite mutual information between the subsystem and the bath. A steady growth of this temperature with the coupling ratio resembles that of the Hawking temperature with the inverse mass of the black hole. Concomitantly, the emergence of non-Fermi liquid behaviour observed near the quantum critical point and in the strong-coupling phase resembles strange-metal-like physics expected near the event horizon from a holographic duality perspective. Our results establish the minimal model as a platform for studying entanglement transfer and information scrambling within a fully unitary quantum framework, and offer new insights into a resolution of the black hole information paradox.
\end{abstract}


\section{Introduction}\label{introsec}

The interplay between quantum information, entanglement dynamics, and emergent thermodynamic behaviour has become a central theme in modern physics, with deep implications ranging from condensed matter systems to quantum gravity. A fundamental question in this context concerns the fate of quantum information when a strongly entangled subsystem interacts with an external environment. Although the evolution of a closed quantum system is governed by unitary dynamics, interactions between subsystems can redistribute quantum correlations and lead to  thermalization, decoherence, and information scrambling  ~\cite{giddings2018quantum, maldacena2020black, li2025simulating, de2024page, chen2020information, wang2024entanglement,su2021page, piroli2020random, liu2021dynamical, pittphilsci27365, hayden2007black, chowdhury2022role, franz2018mimicking, braunstein2013better, kiefer2001hawking, giddings2017nonviolent}. Understanding how entanglement is transferred between subsystems during unitary evolution is therefore of both conceptual and practical importance.
\par
Consider, for example, a quantum system composed of a subsystem and its surrounding environment, where the subsystem is initially prepared in a highly entangled pure state~\cite{almheiri2024universal}. Upon introducing a coupling between the subsystem and the environment, quantum correlations can gradually flow from the subsystem into the environmental degrees of freedom. While the total von Neumann entropy of the complete system remains invariant under unitary evolution~\cite{maldacena2020black}, the entanglement entropy associated with individual subsystems can evolve non-trivially due to the redistribution of correlations. Tracking this transfer of entanglement under many-body quantum dynamics is, however, an inherently difficult problem because of the large number of interacting degrees of freedom involved.
\par
An intriguing possibility is that thermodynamic behaviour is directly emergent from such entanglement transfer processes. In conventional thermodynamics, temperature is defined through the relation \(T = \tfrac{\Delta E}{\Delta S},\) where \(E\) denotes the energy and \(S\) the entropy of the subsystem in equilibrium with a thermal reservoir. From a quantum information perspective, one may instead consider an \emph{effective entanglement temperature} defined through the variation of the ground state energy of the Hamiltonian governing the subsystem ($E$) with respect to the mutual information ($I_{2}$) shared between subsystem and environment, \(T = \tfrac{\Delta E}{\Delta I_2}\)~. 
When the number of degrees of freedom in the environment are substantially larger compared to those within the subsystem, such an information-theoretic description naturally connects entanglement transfer with phenomena such as thermalization, scrambling, and decoherence~\cite{abanin2019}.
\par
These ideas acquire profound significance in the context of the black hole information paradox~\cite{maldacena2020black, piroli2020random, liu2021dynamical, chowdhury2022role, braunstein2013better}. The paradox originates from the apparent incompatibility between quantum mechanical unitarity and black hole evaporation through Hawking radiation. Quantum mechanics requires that a pure state must remain pure under unitary time evolution, thereby ensuring conservation of information. However, Hawking showed that black holes radiate thermally due to particle-antiparticle pair creation near the event horizon~\cite{hawking1975particle}. As the black hole evaporates, the emitted Hawking radiation appears thermal and corresponds to a mixed state. Consequently, if a black hole evaporates completely, an initially pure state describing the ``unevaporated" black hole appears to have evolved into a mixed state describing the emitted radiation, implying information loss and an apparent breakdown of unitarity.
\par
\begin{figure}[!ht]
	\centering
    \includegraphics[scale=0.5]{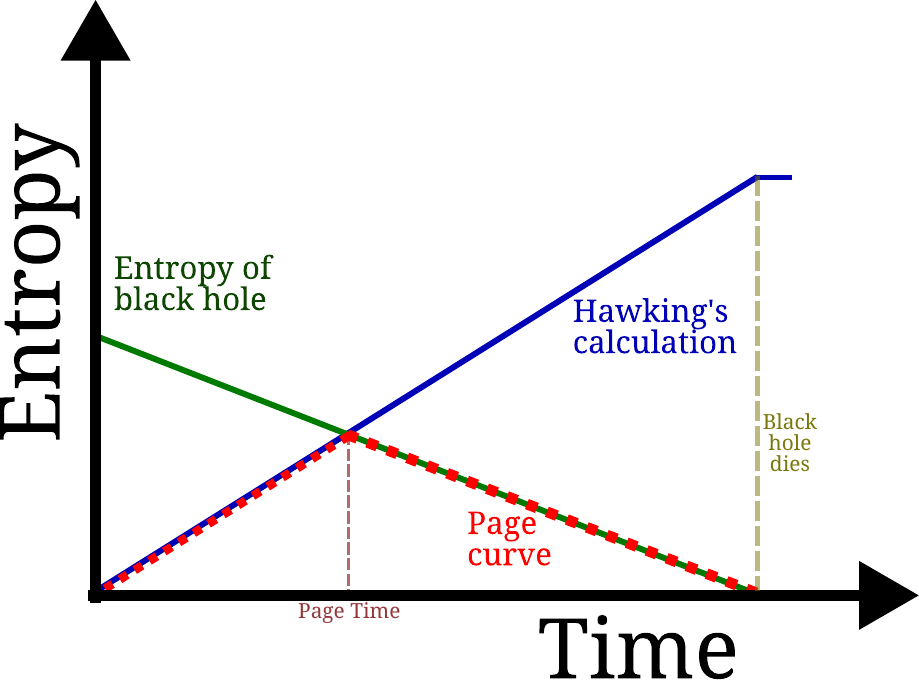}
    \caption{\raggedright Schematic representation of entropy versus time curves during black hole evaporation. The entropy of the black hole (green curve) is expected to fall steadily, while the curve corresponding to Hawking's calculation (blue) corresponds to the monotonic growth of radiation entropy. Finally, the Page curve (red) represents unitary evolution where the radiation entropy eventually decreases after the Page time.}
    \label{fig:HawPageCurve}
\end{figure}

This contradiction can be shown through the entropy evolution during black hole evaporation, shown schematically in Fig.~\ref{fig:HawPageCurve}. During evaporation, the thermodynamic entropy of the black hole decreases with time, while Hawking's semiclassical calculation~\cite{hawking1975particle} predicts that the entropy of the emitted radiation increases monotonically until the black hole  evaporates completely. However, Page argued that unitarity must be preserved if the evaporation process is fundamentally governed by quantum mechanics. In such a unitary evolution, the entropy of the radiation should initially increase, reach a maximum known as the Page time, and subsequently decrease as information is gradually transferred back to the environment. This behaviour, known as the Page curve, has become central to modern discussions of information recovery and quantum scrambling in black hole physics~\cite{maldacena2020black, piroli2020random, liu2021dynamical, chowdhury2022role, braunstein2013better}. Resolving the apparent contradiction between Hawking's prediction and unitary quantum evolution remains one of the most important open problems at the interface of quantum mechanics and gravity. Arguments based on holographic duality~\cite{almheiri2021entropy, almheiri2020replica, penington2022replica, almheiri2019entropy} 
identify nonlocality arising from quantum correlations between regions of the black hole interior and the Hawking radiation on the exterior as key to the restoration of unitarity. However, given the complexity of the approach, the physical origin of these quantum correlations remains an open question.

\par
We explore these questions within the confines of a minimal model for entanglement transfer at $T=0$ that can be described within a unitary framework.
The model consists of a system of two stationary qubits (or spin-\(\tfrac{1}{2}\) magnetic moments) that are strongly entangled to one another (due to an antiferromagnetic Heisenberg exchange coupling \(\tilde{J}\) between them), and with one of the qubits 
also coupled to a bath of itinerant non-interacting electrons through an additional antiferromagnetic exchange interaction ($J_{K}$). 
For a vanishing $J_{K}$, the inter-qubit exchange  interaction $\tilde{J}$ leads to a singlet state between the two qubits. The resulting competition between 
a non-zero \(J_K\) and 
\(\tilde{J}\) is then observed to drive a quantum phase transition (QPT) of the system of qubits,
concomitant with a redistribution of entanglement between the qubits
and the electronic environment. The minimal model provides thereby a natural platform by which to investigate entanglement transfer in a strongly correlated many-body system. 
\par
We note that quantum mechanical models have been employed earlier towards understanding how information can leave an evaporating black hole. For instance, in Refs.\cite{mathur2009information,mathur2011information,mathur2011correlations}, each emission step creates a Hawking pair of entangled qubits, one escaping as radiation and the other falling inside the black hole. Strong subadditivity then shows that small corrections to the evolution cannot reduce the growing entanglement of the radiation, indicating that substantial modifications are needed to restore unitarity. In Refs.\cite{giddings2013nonviolent,giddings2013quantum}, quantum mechanical and field-theory approaches have been employed in describing the transfer of information between degrees of freedom in the interior, the near-horizon and the environment just exterior to the black hole. Ref.\cite{hayden2007black,liu2021dynamical,lau2022} treat the state of the black hole as obtained from the action of a random unitary operation on a system of qubits. In several of these works, the entanglement dynamics between the black hole and environment are generated by a fixed unitary map. We take instead a complementary approach based on a minimal Hamiltonian model for the interior, boundary and environment qubits (see Fig.\ref{figBHdm}): the entanglement transfer and associated quantum dynamics arise naturally from the interplay of various couplings between the qubits, and can be followed explicitly within a unitary framework. This likely offers insight into the physical origins of the nonlocal quantum correlations obtained from holographic duality calculations ~\cite{almheiri2021entropy, almheiri2020replica, penington2022replica, almheiri2019entropy} 
within a microscopic (albeit toy model) approach.
\par
The qubit models studied in Refs.\cite{kehrein2024,saha2024} also deserve mention in this context: here, a gas of fermions confined initially to a finite region is then allowed to scatter into a semi-infinite region that is initially empty. The changes in the entanglement structure with time are observed to resemble the Page curve that is associated with a quantum phase transition in the entanglement Hamiltonian. By contrast, the evolution of the entanglement in our model arises from the many-body interactions between the finite-sized quantum spin system and a fermionic bath, and displays a quantum phase transition of the many-particle Hamiltonian and its eigenspectrum. Our approach is similar to that taken in Ref.\cite{chen2020}, where the Page curve arising from the entanglement dynamics in a SYK model coupled to a bath of non-interacting Majorana fermions was studied via the Schwinger-Dyson equations in the large-$N$ limit. In Refs.\cite{dhar2025} and \cite{jha2025}, the Page-like curve was obtained for a finite-sized XXZ model coupled to a semi-infinite XXZ chain whose initial state corresponds to an empty sea of spinless electrons and a non-interacting gas of spinless electrons respectively. The model adopted by us is simpler than those in Refs.\cite{chen2020,dhar2025,jha2025}, and amenable to analysis without the need for taking either a large $N$ or any other semiclassical limit.
\par
We begin by first describing the minimal model in Section~\ref{model}. To access these phenomena, it is essential to probe the low-energy sector of the theory. For this purpose, in Section \ref{urgsection}, we employ the non-perturbative Unitary Renormalization Group (URG) method~\cite{mukherjee2020holographic-A, mukherjee2020holographic-B}. The method has previously been successfully applied to a wide variety of correlated quantum systems~\cite{mukherjee2020scaling-A, mukherjee2020scaling-B, pal2019correlated, patra2021origin, mukherjee2021fermionic, mukherjee2022unveiling, mukherjee2023kondo, debata2026kondo}. For a detailed description of the formalism, we refer the reader to the Methods subsection of Ref.~\cite{debata2026kondo}. Our analysis reveals a quantum phase transition of the qubit-subsystem as a function of $J_{K}/\tilde{J}$ that is concomitant with a transition in the nature of the entanglement within the ground state. This transition is discussed as being analogous to a continuous Hawking-Page transition of a black hole \cite{hawking1983,gursoy2010,gursoy2011}. In Section \ref{entsection}, we demonstrate that the entanglement dynamics of the minimal
model exhibit several remarkable features analogous to the physics of black hole evaporation. In particular, the behaviour of certain bipartite and tripartite mutual information defined among qubit and bath degrees of freedom resembles, as a function of the coupling ratio \(J_K/\tilde{J}\), the qualitative structure of the entropy evolution during black hole evaporation as described in Fig.\ref{fig:HawPageCurve} above. The emergence of negative tripartite mutual information indicates multipartite entanglement structures strongly reminiscent of information scrambling observed in systems that obey the holographic duality~\cite{hosur2016chaos, hayden2013holographic}, and a non-monotonic nature that resembles qualitatively the Page curve. Furthermore, the effective entanglement temperature defined in Section \ref{tempsection} exhibits similarities with the Hawking temperature. In addition, we find the emergence of non-Fermi liquid behaviour at and near the quantum critical point (see Section \ref{nflsection}), drawing parallels with physics near the event horizon observed from calculations involving the holographic duality~\cite{phillips2022stranger,sachdev2015bekenstein,faulkner2010black}. In Section \ref{scramsection}, we present results for the time dynamics of an out-of-order-correlator (OTOC) for the degrees of freedom representing the qubits that suggest scrambling behaviour. This is reinforced by the temporal evolution of several entanglement measures that display behaviour analogous to entropy evolution during black hole evaporation. We conclude in Section \ref{concsection}, and present technical details of some calculations in the appendices. For the ease of the reader, we present below a summary of the major results obtained by us.\\



\noindent
\textbf{Summary of major results :}
\begin{itemize}
    \item Competition between the coupling within the subsystem and the coupling to the environment drives a quantum phase transition (see Fig.\ref{figBHdm}). On one side of the transition, the two qubits form a singlet and decouple from the bath. We denote this as the weak coupling phase from the perspective of the system-bath interaction. On the other side, the boundary qubit is entangled with the bath and the interior qubit is weakly coupled to the boundary; this is denoted as the strong coupling phase of the system-bath interaction. The former is analogous to the ``formation'' phase of a black hole, and the latter to its ``evaporation'', i.e., the analog of a Hawking-Page transition.

    \item In the strong coupling phase, the mutual information between the interior and boundary qubits decays, while that between the boundary qubit and the environment grows. This is analogous to the fall and rise of black hole entropy and radiation entropy respectively. Further, the tripartite mutual information between the two qubits and the environment is negative and non-monotonic, resembling a Page curve and indicating scrambling.

    \item A temperature defined from the energy change per unit change of subsystem - bath mutual information grows linearly in the evaporation phase, analogous to the increase in Hawking temperature with decreasing black hole mass.

    \item Near the transition and in the strong-coupling phase, the system entanglement is observed to undergo a transfer into that between a subsystem and the environment via the emergence of low-lying gapless excitations that are non-Fermi liquid in nature. This is seen from the effective Hamiltonian for the excitations, as well as a pseudogapped spectral function and a divergent self-energy obtained from a local Greens function for the boundary qubit. This parallels the observation of non-Fermi liquid excitations lying proximate to the horizon in quantum models that are holographic duals of a black hole.

   	\item An out-of-order-correlator (OTOC) shows exponential growth in the strong coupling phase, with a Lyapunov exponent that increases with the coupling to the bath. Various real-time entanglement measures indicate thermalisation, i.e., they mimic black hole entropy evolution deep inside the evaporation phase.
\end{itemize}


\section{A Minimal Two-qubit Model for Entanglement Transfer}\label{model}
We formulate a simple model of two qubits (spin-$1/2$ moments $S_{d}$ and $S_{d'}$) that are coupled to one another through an antiferromagnetic Heisenberg exchange coupling ($\tilde{J}>0$), and one of them $S_{d}$ coupled to to the local spin-moment ($S_{0}$) of a bath of non-interacting conduction electrons through another antiferromagnetic Heisenberg exchange interaction ($J_{K}>0$, also known as the Kondo interaction~ \cite{kondo1964resistance,anderson1969exact,anderson1970exact,anderson1970poor,bulla2008numerical,wilson1975renormalization,mukherjee2022unveiling,hewson1997kondo}): 
\begin{equation}
H = \sum_{k,\sigma} \varepsilon_k n_{k\sigma}
+ J_K S_d \cdot S_0
+ \tilde{J} S_{d'} \cdot S_d~,
\label{bareHam}
\end{equation}
where the first term represents the kinetic energy of the conduction electrons.
We recall that the interaction mediated by the coupling $J_{K}$ is equivalent to the venerable spin-boson problem~\cite{caldeira1983quantum,chakravarty1984dynamics,guinea1985bosonization,leggettRMP1987}, i.e., a dissipative coupling of a qubit with a reservoir of bosonic oscillators (whose spectral function is of Ohmic form) that can lead to an irreversible damping of the quantum dynamics of the qubit. The ratio $J_K/\tilde{J}$ represents a tuning parameter that controls the degree of coupling of the $d$-qubit to the environment vis-a-vis coupling to the $d'$-qubit (i.e., within the subsystem). For $J_{K}=0$, the two qubits are decoupled altogether from the conduction bath and form a singlet ground state due to the coupling $\tilde{J}$. On the other hand, for $\tilde{J}=0$, the coupling $J_{K}$ leads to the formation of a singlet between $S_{d}$ and the qubit $S_{0}$ (which represents the electronic screening cloud at the location of the $d$-qubit) of the bath, and the qubit $S_{d'}$ decouples completely from the other qubit ($S_{d}$). The competition between the inter-qubit interaction $\tilde{J}$ and the (Kondo) screening due to $J_{K}$ thus defines a transfer of the entanglement among the three parties (i.e., the three qubits $S_{d}, S_{d'}$ and $S_{0}$) in passage between these two maximally entangled singlet states.
\par
As we will see below, this prototypical model captures key ingredients relevant to discussions of entanglement transfer present in the physics of black hole evaporation~ \cite{giddings2013,giddings2018quantum,giddings2019black,braunstein2013better}. In this analogy, the three qubits $S_{d'}$, $S_{d}$ and $S_{0}$ represent fermionic degrees of freedom inside, at the boundary and the near-boundary environment of the black hole respectively~\cite{giddings2013}. Thus, the interaction term $S_{d'}\cdot S_{d}$ represents a quasi-local interaction between the black hole degrees of freedom at the interior and boundary. Similarly, the interaction term $S_{d}\cdot S_{0}$ represents quasi-local interactions between the black hole degree of freedom at the boundary and a fermionic degree of freedom in the nearby environment.
Finally, the first term of the Hamiltonian (eq.\eqref{bareHam}) represents the kinetic energy of the electronic environment. 
Consistent with the framework proposed for an open quantum system in Refs.\cite{giddings2013,giddings2018quantum,giddings2019black}, the Hilbert space dimension of the environment is much larger than that of the qubit subsystem within this minimal model.


\section{Entanglement Transfer Across a Quantum Phase Transition}\label{urgsection}
In order to understand better the implications of the competition between the frustrating interactions $\tilde{J}$ and $J_{K}$ in the minimal model and identify the low-energy phases, we perform a  Unitary Renormalization Group (URG) analysis of the 
Hamiltonian eq.\eqref{bareHam}. This analysis obtains the renormalization group (RG) flow equations for various couplings by integrating out the high energy degrees of freedom of the conduction bath by the iterative application of many-particle unitary transformations~\cite{mukherjee2020holographic-A, mukherjee2020holographic-B}. As this method has already been applied to a variety of problems~\cite{mukherjee2020scaling-A, mukherjee2020scaling-B, pal2019correlated, patra2021origin, mukherjee2021fermionic, mukherjee2022unveiling, mukherjee2023kondo, debata2026kondo}, we postpone all details of the calculation in Appendix~\ref{appURGDetail} and present only the result here. 
\par
The renormalization group (RG) flow equation for the antiferromagnetic coupling $J_K$ is obtained as:
\begin{equation}
    \frac{\Delta J_K}{\Delta D} = -\frac{J_K^2}{2} \left( \frac{1}{\omega - E_1} + \frac{1}{\omega - E_2} \right) N(D)~,
    \label{kondorg}
\end{equation}
where the energies $E_{1}$ and $E_{2}$ are given by
\begin{equation}
    E_1 = \frac{D}{2} - \frac{J_K}{4} + \frac{\tilde{J}}{4}~~~~, \quad
    E_2 = \frac{D}{2} - \frac{J_K}{4} - \frac{\tilde{J}}{4}~,
\end{equation}
where $\omega$ represents the energy scale for the quantum fluctuations being intergrated out and $D$ is the bandwidth of the conduction bath kinetic energy. The inter-qubit coupling $\tilde{J}$ remains marginal under the URG flow. The RG equation eq.\eqref{kondorg} indicates a critical point determined by the condition
\begin{equation}
    \omega - \frac{D}{2} + \frac{J_K}{4} + \frac{\tilde{J}}{4} = 0~.
\end{equation}
For representative parameters $D=1$ (i.e., taken in units of the hopping amplitude), $\omega = -D/2$ and the bare values of the couplings $J_K=2=\tilde{J}$, this condition yields a critical ratio $(J_K/\tilde{J})_{C} = 1$.

\begin{figure}[!ht]
\centering
\fbox{\includegraphics[scale=0.52]{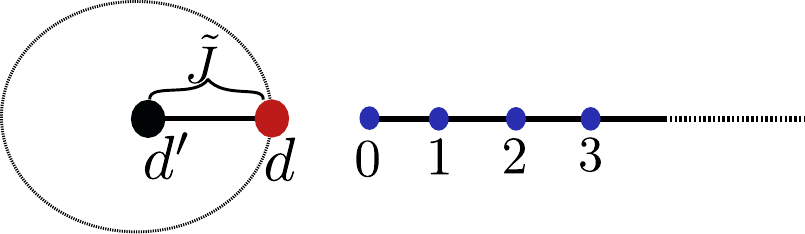}}
\fbox{\includegraphics[scale=0.5]{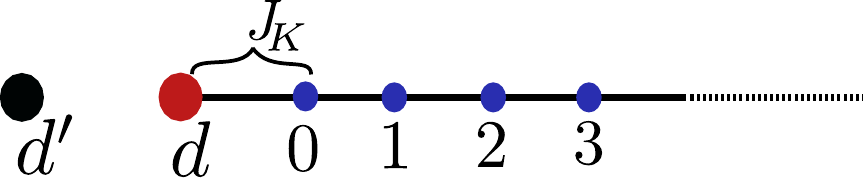}}
\caption{\raggedright (Left) Schematic diagram of singlet between the two qubits ($d'$ and $d$) via the coupling $\tilde{J}$ in the weak-coupling phase, while the conduction bath decouples. (Right) Schematic diagram of the singlet formed between the $d$-qubit and the bath $0$th site in the strong-coupling phase, and with an (almost) decoupled $d'$-qubit.}
\label{figBHdm}
\end{figure}

For the ratio of bare parameters $J_K/\tilde{J} <  (J_K/\tilde{J})_{C} = 1$, the RG flow drives the coupling $J_{K}$ to its weak-coupling fixed point at $J_K^* = 0$. As mentioned earlier, this results in the formation of a singlet between the two qubits ($S_{d'}$ and $S_{d}$) via the coupling $\tilde{J}$, while the conduction bath decouples. A schematic representation of this is shown in Fig.~\ref{figBHdm}~(Left). In this regime, the effective Hamiltonian is given by
\begin{equation}
    H_{\mathrm{eff}} = \sum_{k,\sigma} \varepsilon_k\, n_{k\sigma} + \tilde{J}\, \mathbf{S}_{d'} \cdot \mathbf{S}_d~.
\end{equation}

\par
In contrast, for the ratio of bare parameters $J_K/\tilde{J} > (J_K/\tilde{J})_{C} = 1$, the RG flow enhances $J_{K}$ to a strong-coupling fixed point value $J_K^*$. This leads in turn to the formation of a singlet between the qubits $S_{d}$ $S_{0}$ (i.e., the local site of the conduction bath that couples to the $d$-qubit). In the strong-coupling limit $J_K^* \gg \tilde{J}$, the qubit $S_{d'}$ effectively decouples from the rest of the system. A schematic representation of this is shown in Fig.~\ref{figBHdm}~(Right).  Thus, the critical ratio $(J_K/\tilde{J})_{C} = 1$ represents a quantum phase transition (QPT) for the two-spin system involving a dramatic change in the nature of the many-body ground state. The corresponding effective Hamiltonian for this strong-coupling limit is given by
\begin{equation}
    H_{\mathrm{eff}} = \sum_{k,\sigma} \varepsilon_k\, n_{k\sigma} + J_K \mathbf{S}_d \cdot \mathbf{S}_0~.
\end{equation}
We note in passing that for the case of a bare ferromagnetic coupling between the $d$-qubit and the bath ($J_{K}<0$), this interaction with the environment is irrelevant and vanishes at the fixed point $J_{K}^{*}=0$~ \cite{kondo1964resistance,anderson1969exact,anderson1970exact,anderson1970poor,bulla2008numerical,wilson1975renormalization,mukherjee2022unveiling,hewson1997kondo}. Thus, only the weak-coupling phase (see Fig.~\ref{figBHdm}~(Left)) can be realised here. It is worth noting that the QPT is analogous to the Hawking-Page transition at finite temperature from a black hole phase to one in which it is replaced by a thermal state \cite{hawking1983}. While the original work of Hawking and Page obtained a first order transition, we obtain a Berezinskii-Kosterlitz-Thouless \cite{berezinskii1971,kosterlitz1973} type continuous QPT reminiscent of the proposal of Refs.\cite{gursoy2010,gursoy2011}. This QPT involves the confinement-deconfinement transition for topological excitations corresponding to the spin-flip processes of the conduction bath electrons that (Kondo) screen the spin-1/2 impurity $d$ to which they are locally coupled~\cite{anderson1970exact}. We recall that, employing the idea of holographic duality, Witten demonstrated that the Hawking-Page transition can be mapped onto a confinement-deconfinement transition in a quantum gauge theory \cite{witten1998}. It is, therefore, interesting to speculate that the QPT observed by us can similarly be mapped onto a Hawking-Page transition.
\begin{figure}[!ht]
\centering
\fbox{\includegraphics[scale=0.5]{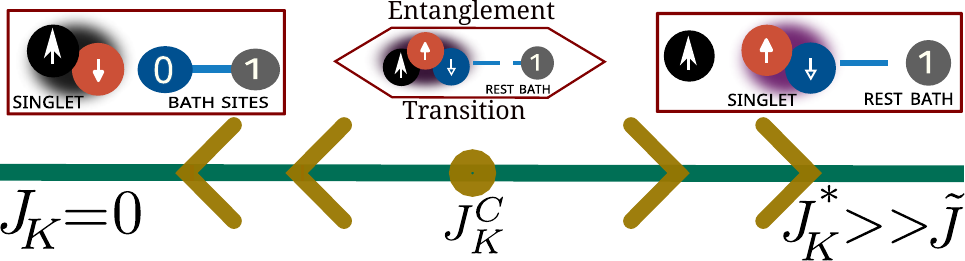}}
\caption{\raggedright Schematic diagram of entanglement transfer across the quantum phase transition in the minimal model. For $J_K < J^c_K$, the entanglement resides within the qubit system $S_d$ and $S_d^\prime$, whereas for $J_K > J_K^c$, the entanglement is transferred to the boundary + environment subsystem comprised of the $S_d$ and $S_0$ qubits. In between, the state at the transition displays tripartite information between the three-party composite system of $S_d$, $S_d^\prime$ and $S_0$.}
\label{figEntTransfer}
\end{figure}

\begin{figure}[!ht]
    \centering
    \includegraphics[scale=0.3]{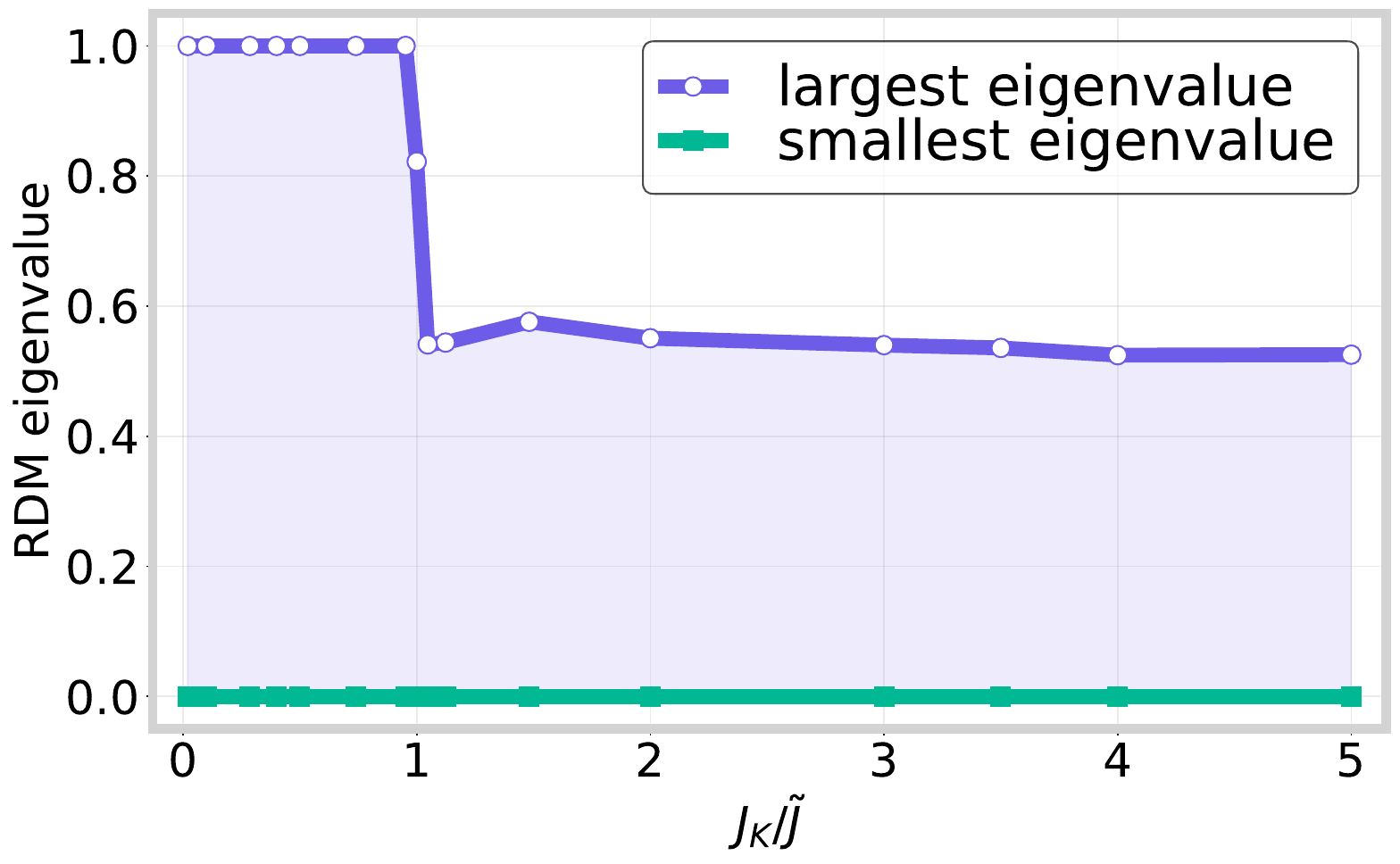}
    \caption{Variation of largest (purple curve) and smallest (green curve) eigenvalue of reduced density matrix of $d'-d$ bipartition with varying coupling ratio $J_K/\tilde{J}$. The sudden jump in the purple curve across the critical ratio $J_K/\tilde{J}_{C}=1$ demonstrates the entanglement transition concomitant with the quantum phase transition.}
    \label{fig:RDMeigval}
\end{figure}

We now show that the QPT observed above coincides with a transition in the entanglement spectrum of our quantum system. For this, we  investigate the eigenvalue spectrum of the reduced density matrix for the $d'-d$ bipartition. As shown in Fig.~\ref{fig:RDMeigval}, the largest eigenvalue remains unity for $J_K/\tilde{J}<1$, whereas it decreases abruptly to a value between $0.5$ and $0.6$ beyond the transition. This pronounced change
indicates a substantial reconstruction of the reduced density matrix across the quantum phase transition. In contrast, the smallest eigenvalue remains zero throughout the parameter regime considered. The spectrum of the reduced density matrix also provides a direct connection to the entanglement (or modular) Hamiltonian, $H_E$, defined as
\begin{equation}
\rho_A = \frac{e^{-H_E}}{Z_A}~,
\qquad
Z_A = \mathrm{Tr}\left(e^{-H_E}\right)~.
\end{equation}
For $\lambda_i$ denoting an eigenvalue of $\rho_A$, the corresponding entanglement energy is given by $\xi_i = -\ln(\lambda_i Z_A).$
In particular, the largest eigenvalue of the reduced density matrix determines the ground-state entanglement energy. Therefore, the observed change in the 
entanglement spectrum acts as a complementary probe of the quantum phase transition. 
Interestingly, a reconstruction of the reduced density matrix spectrum was observed in Ref.~\cite{kehrein2024} to precede 
the Page-curve-like entanglement dynamics in a model of non-interacting electrons in a model of non-interacting electrons. In sections \ref{entsection} and \ref{scramsection} respectively, we demonstrate entanglement transfer in the strong coupling phase that is analogous to the Page curve in evolution under renormalisation as well as temporal dynamics. 

\par\noindent
\textbf{\textit{Entanglement transfer and an analogy to black-hole physics:}}~At the quantum critical point corresponding to $(J_{K}/\tilde{J})_{C}=1$, all three qubit degrees of freedom ($d'$, $d$, and the bath site $0$) compete to form singlets, signaling a nontrivial redistribution of entanglement where neither of the singlets of the weak- and strong-coupling phases can be formed. 
In keeping with our earlier discussion, this redistribution of entanglement admits an analogy to black-hole physics. 
In the regime $J_K^* = 0$, the $d'-d$ singlet is formed under the RG flow and effectively decouples from the environment. This resembles a phase in which the black-hole is formed with degrees of freedom in its interior and boundary coupled to one another. On the other hand, upon increasing $J_K$ beyond the critical point, the environment becomes increasingly entangled with the $d$-qubit, leading to a decoupling of the $d'$-qubit from the rest. This is analogous to an evaporation phase of the black-hole by a decoherence of its boundary, leaving exposed its interior. At the critical point itself, all three qubits representing the black interior, its boundary and the near-boundary environment are strongly coupled and must mediate the entanglement transfer~ \cite{giddings2013,giddings2018quantum,giddings2019black, braunstein2013better}. We turn to this next.

\section{Entanglement Properties}\label{entsection}
\begin{figure}[!ht]
	\centering
    \includegraphics[scale=0.32]{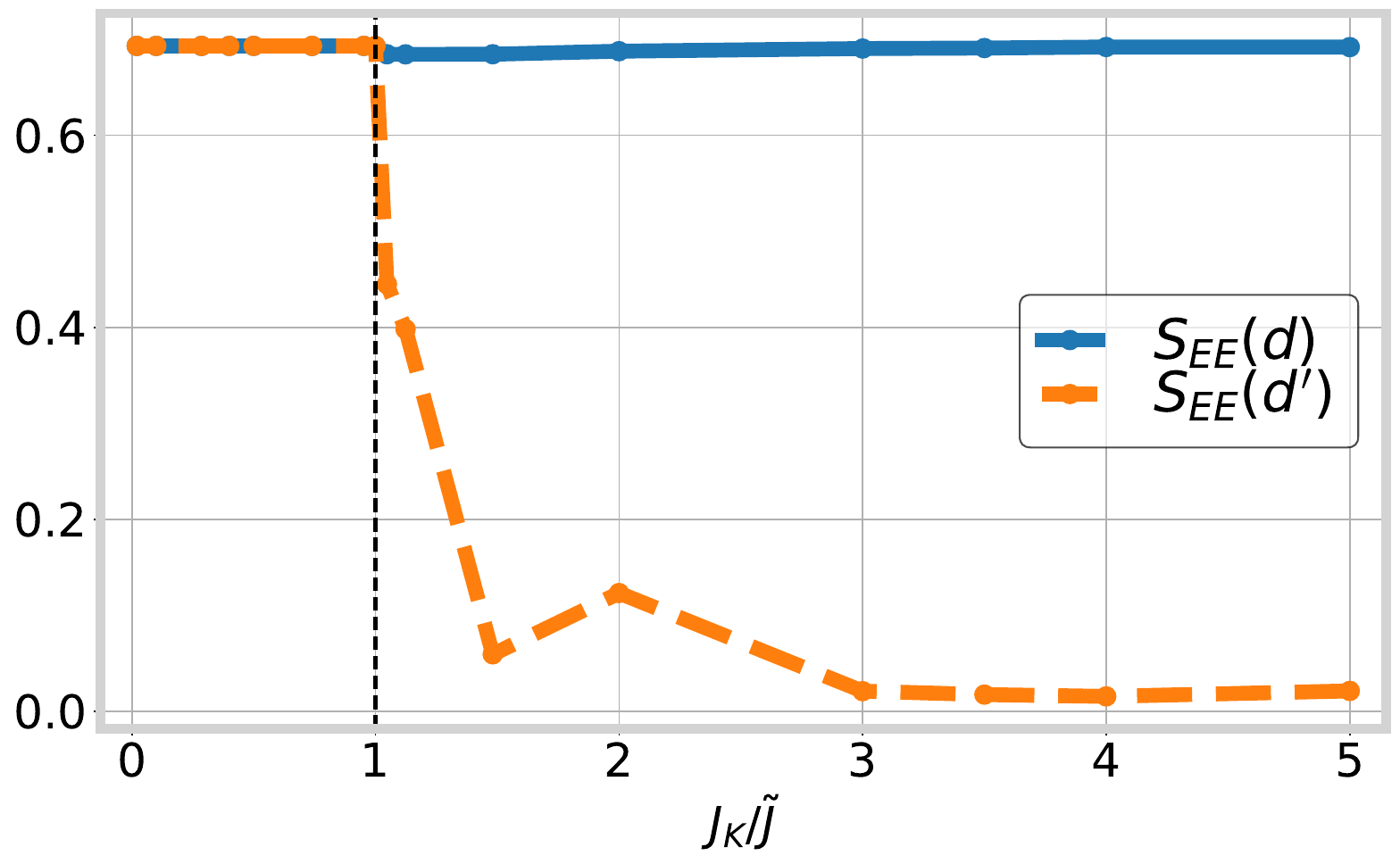}
    \caption{\raggedright Variation of entanglement entropy of the qubits \(d\) and \(d'\) with varying coupling ratio $J_{K}/\tilde{J}$. See main text for discussion.}
    \label{fig:EntEntropy}
\end{figure}

\begin{figure}
	\centering
    \includegraphics[scale=0.28]{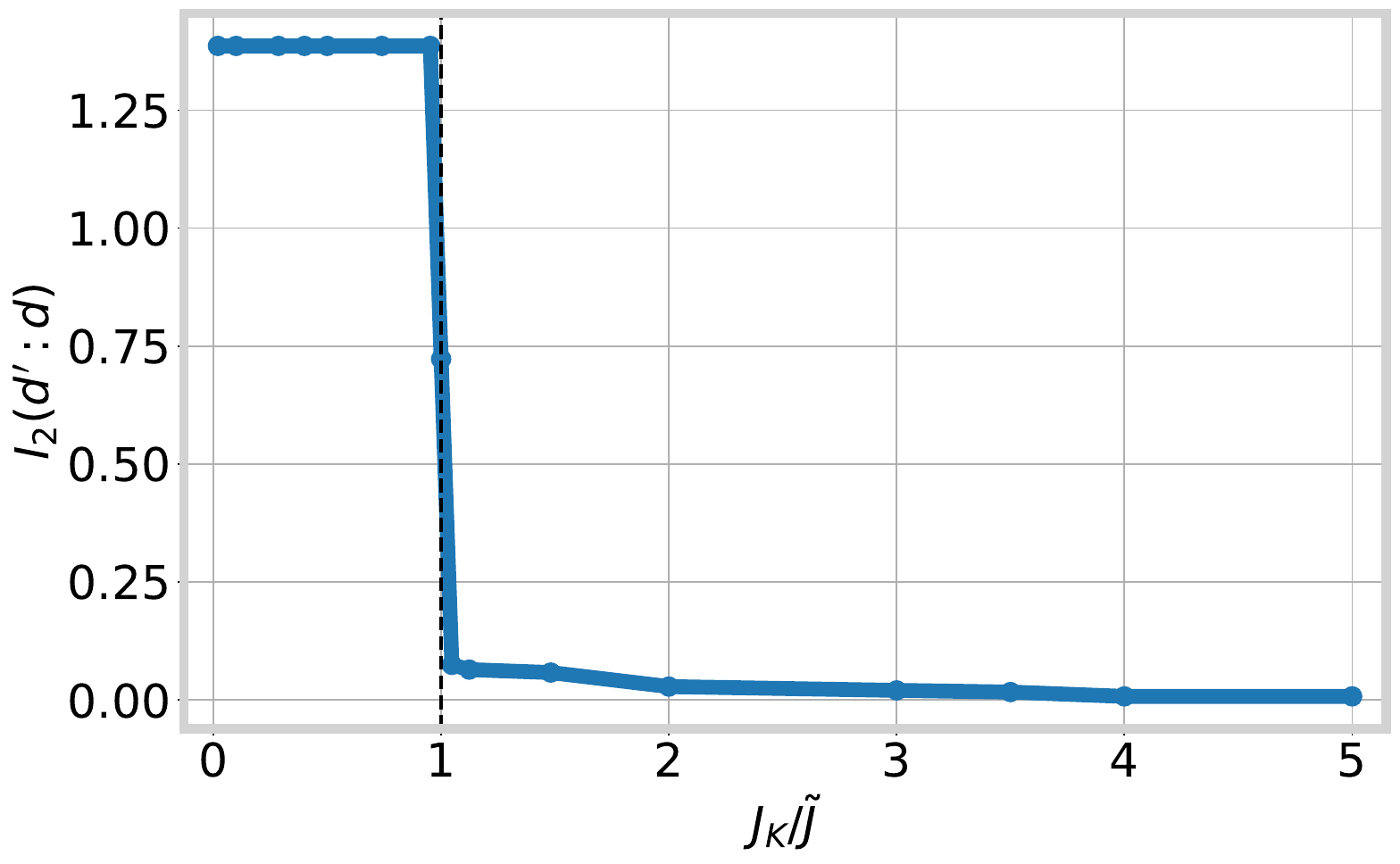}
    \includegraphics[scale=0.28]{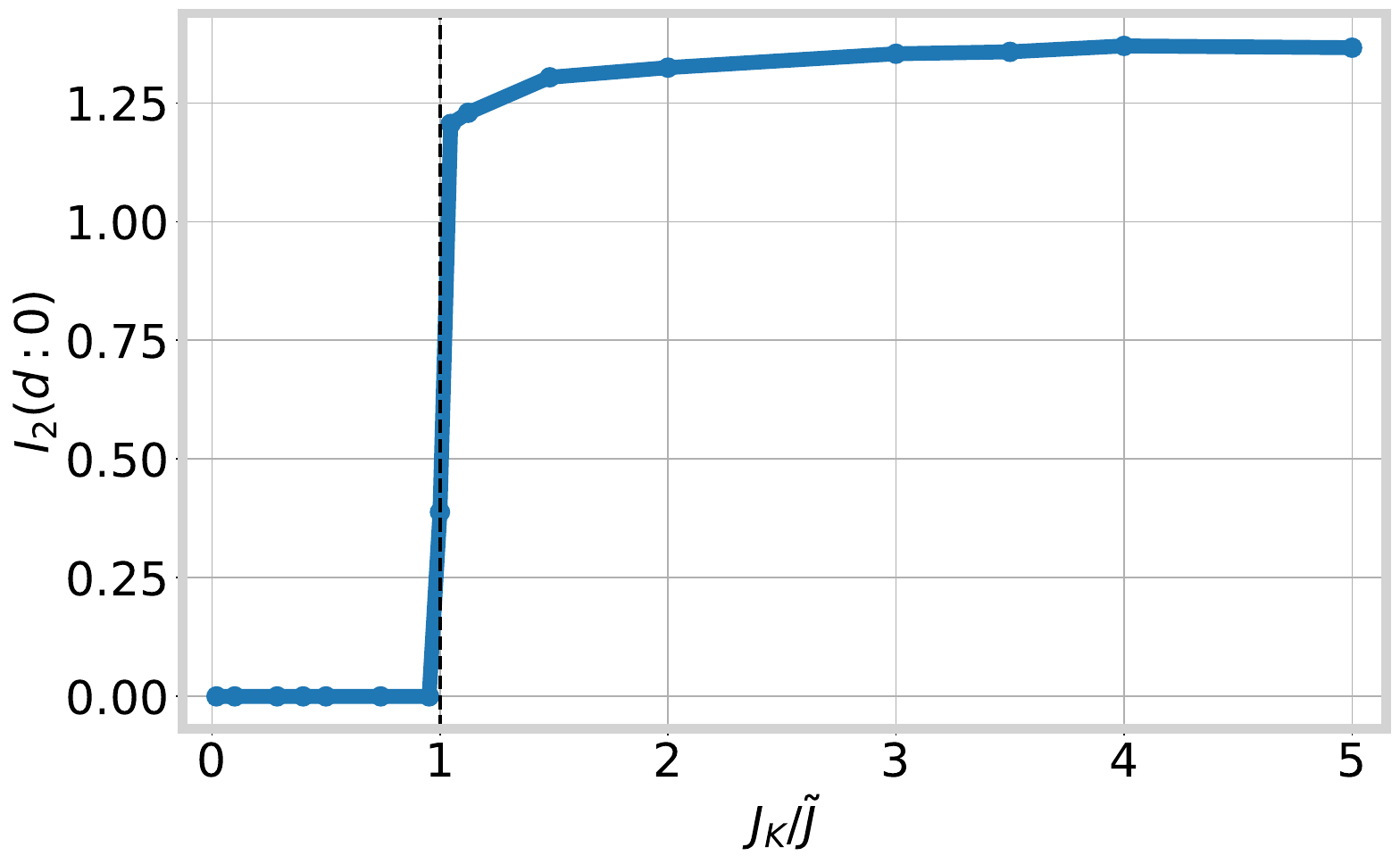}
    \caption{\raggedright Variation of (Left) Bipartite mutual information between $d'$ and $d$ ($I_2(d':d)$) and (Right) Bipartite mutual information between $d$ and the $0$th bath site ($I_2(d:0)$) with varying coupling ratio $J_{K}/\tilde{J}$. See main text for discussion.}
    \label{fig:I2dprimedI2d0}
\end{figure}

\begin{figure}
	\centering
    \includegraphics[scale=0.28]{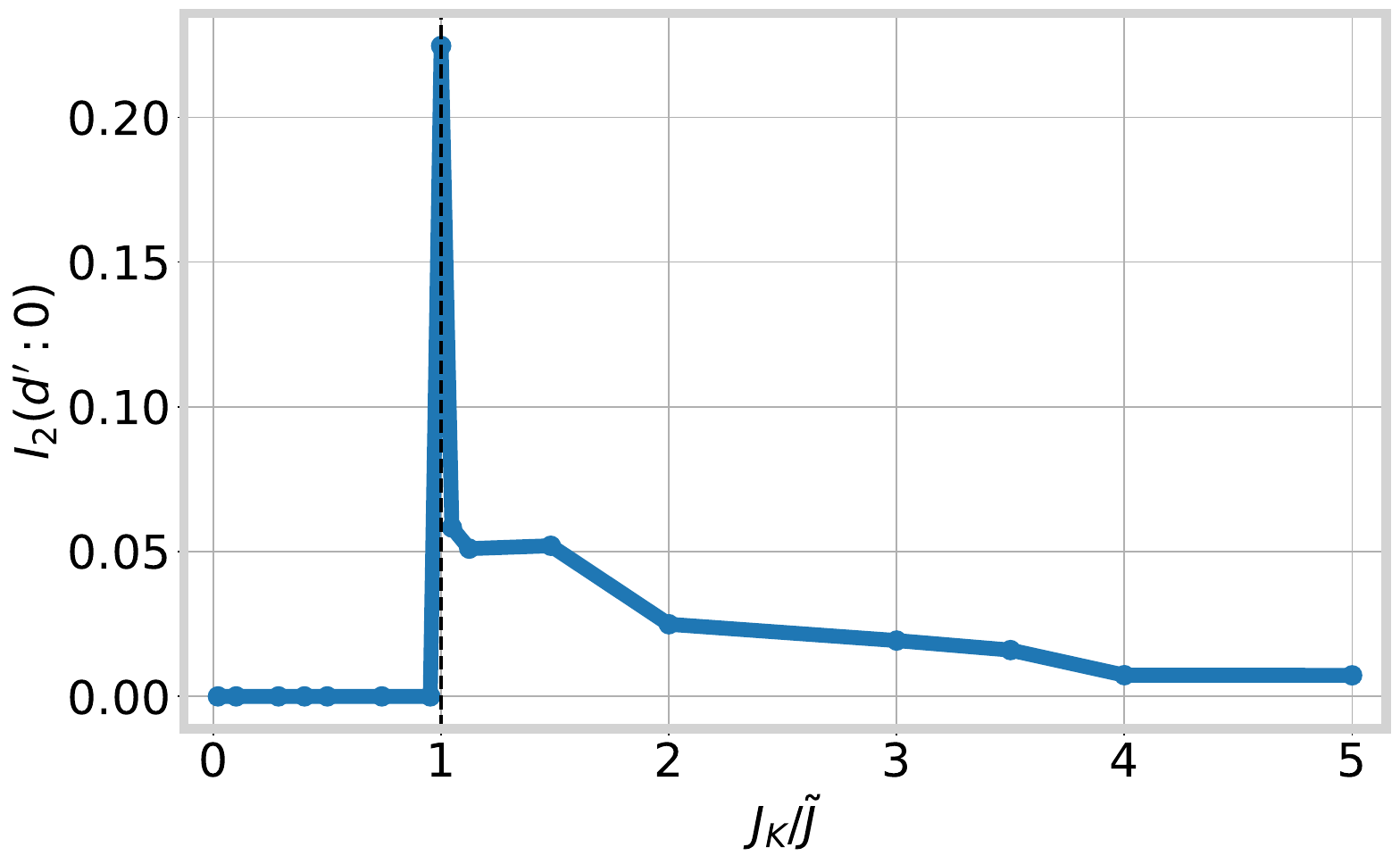}
    \includegraphics[scale=0.28]{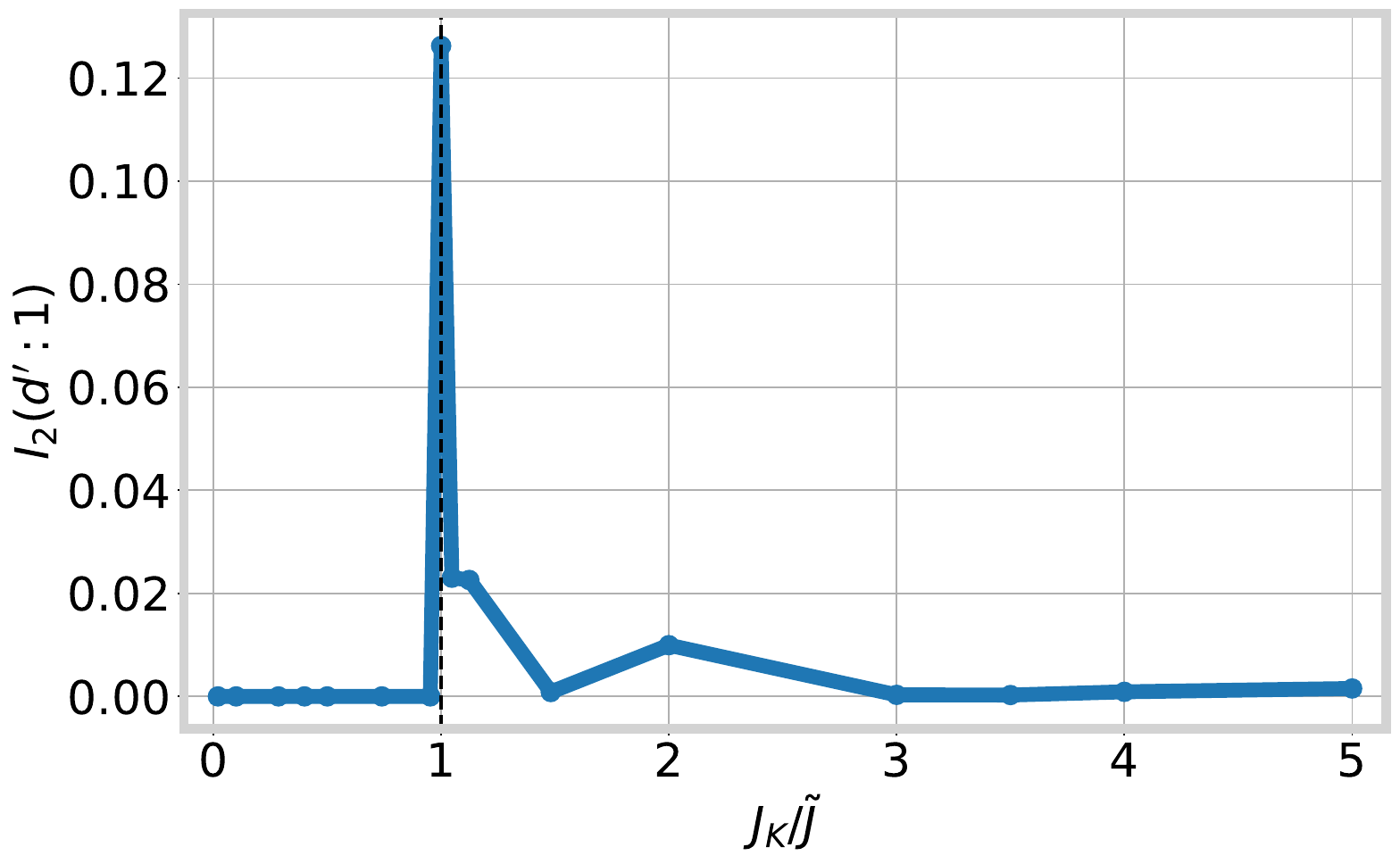}
    \caption{\raggedright Variation of (Left) Bipartite mutual information between $d'$ and the $0$th bath site $I_2(d':0)$) and (Right) Bipartite mutual information between $d'$ and the $1$st bath site ($I_2(d':1)$) with varying coupling ratio $J_{K}/\tilde{J}$. See main text for discussion.}
    \label{fig:I2dprime0I2dprime1}
\end{figure}

\begin{figure} 
\centering
\includegraphics[scale=0.32]{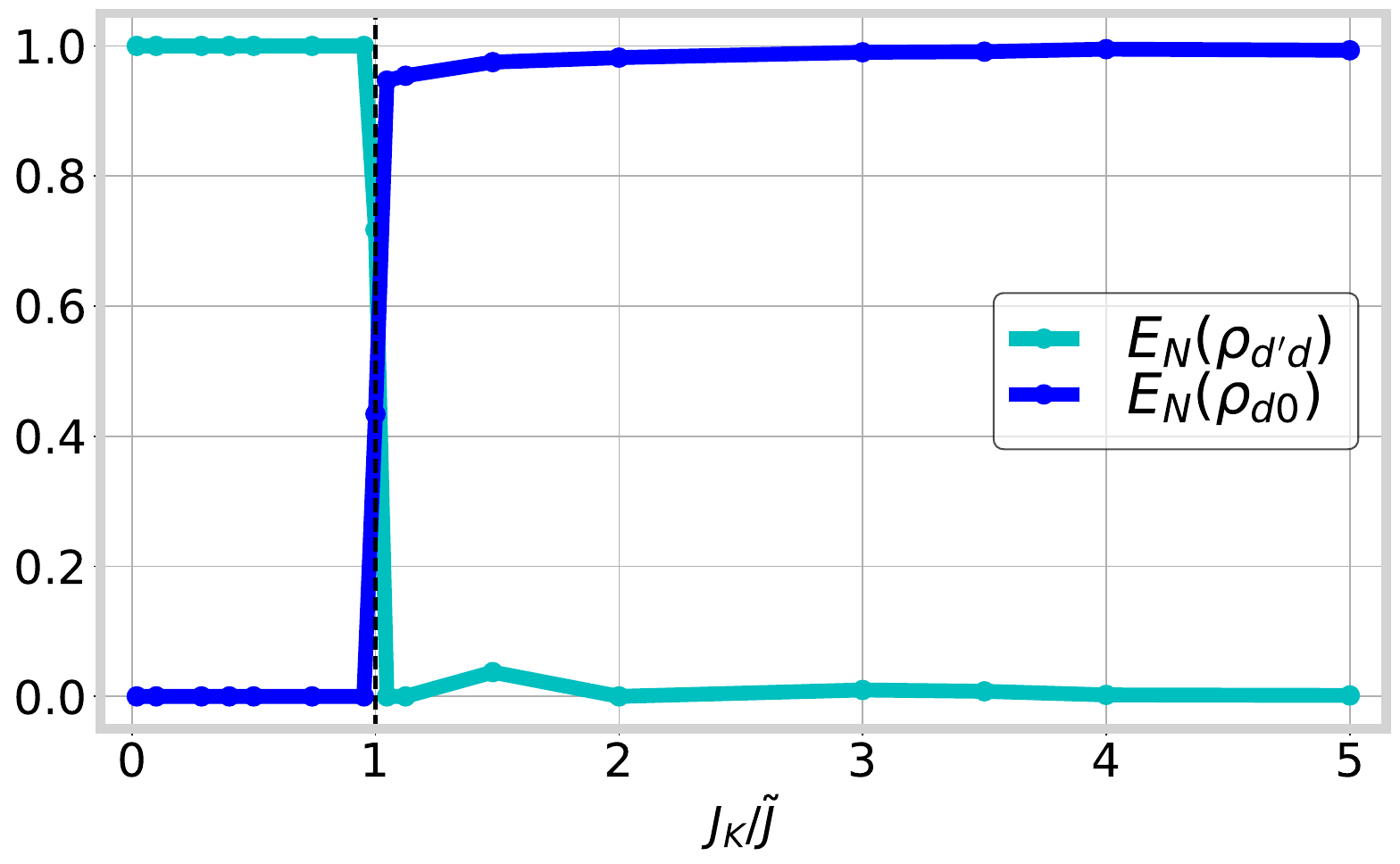} 
\caption{\raggedright Variation of the logarithmic negativity for $\rho_{d^\prime d}$ ($E_N(\rho_{d^\prime d})$) and $\rho_{d0}$ ($E_N(\rho_{d0})$) with varying coupling ratio $J_K/\tilde{J}$. See main text for discussion.} 
\label{fig:LogNegativity} 
\end{figure}

\begin{figure}
	\centering
    \includegraphics[scale=0.32]{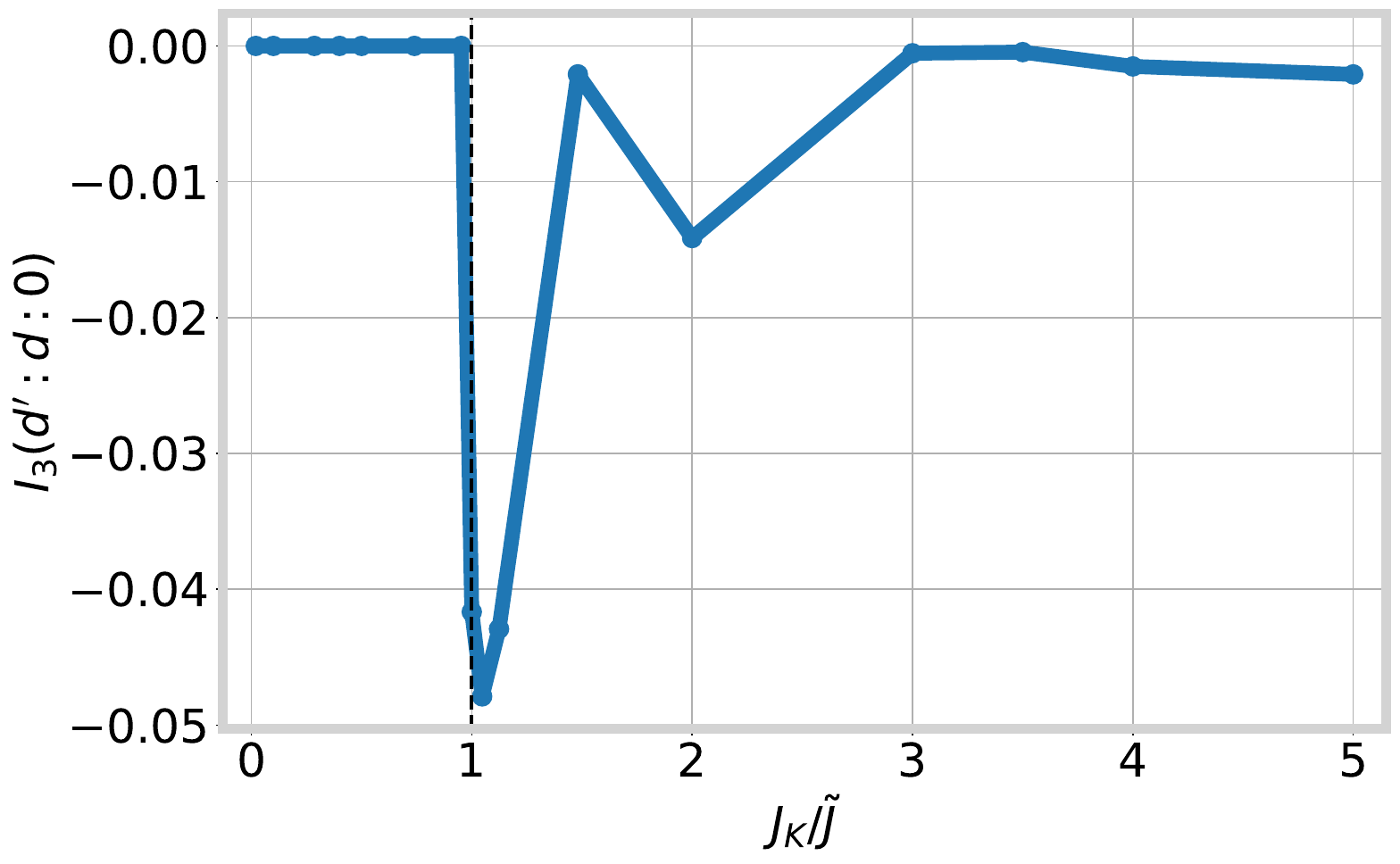}
    \caption{\raggedright Variation of tripartite mutual information ($I_3(d':d:0)$) with varying coupling ratio $J_{K}/\tilde{J}$. See main text for discussion.}
    \label{fig:Tripartite}
\end{figure}

In order to investigate the redistribution of entanglement among various subsytems, we now quantify the entanglement transfer process across the quantum phase transition through the lens of several measures of entanglement, including the entanglement entropy of the $d'$- and $d$-qubits, the bipartite ($I_{2}$) and tripartite ($I_{3}$) mutual informations  among the three qubits $(d',d,0)$ and the logarithmic negativities ($E_N$) of the reduced density matrices corresponding to the $d'-d$ and $d-0$ bipartitions. These quantities are computed from the fixed point theories obtained from the unitary RG analysis explained above: Figs.~\ref{fig:EntEntropy}-\ref{fig:Tripartite} below represent their evolution under renormalisation, with  the x-axes corresponding to the bare value of the coupling ratio $J_{K}/\tilde{J}$.\\
\par\noindent
\textbf{\textit{Entanglement Entropy:}}~We see in Fig.~\ref{fig:EntEntropy} that the entanglement entropy of the \(d^{\prime}\)-qubit (\(S_{EE}(d^{\prime})\)) assumes the maximum value of $\ln 2$ before the critical point due to the strong $d'$-$d$ qubit coupling \(\tilde{J}\), but decreases beyond the transition and approaches zero as the $\tilde{J}$ weakens in comparison to $J_{K}$. In contrast, the entanglement entropy of the $d$-qubit (\(S_{EE}(d)\)) remains nearly constant at \(\log 2\), with only a slight dip in value very near the transition ($J_{K}/\tilde{J}\sim 1+$). Remarkably, this indicates that the $d$-qubit is (almost) maximally entangled on both sides of the quantum phase transition. 

\par
In terms of a comparison of this to the physics of black hole evaporation, the behaviour of \(S_{EE}(d)\) can be interpreted as an analogue of the black hole event boundary acting as a \emph{bridge} for the quantum information between the black hole interior and its surrounding environment. In the black hole formation phase ($(J_{K}/\tilde{J}) < 1$), the event horizon predominantly shares information with the black hole interiori through strong inter-qubit coupling $\tilde{J}$. In contrast, during the evaporation phase ($(J_{K}/\tilde{J}) > 1$), the \(d\)-qubit is increasingly entangled with the fermionic environment through the exchange oupling \(J_K\). 
The behaviour of \(S_{EE}(d^{\prime})\) further supports this picture: the maximum value of \(S_{EE}(d^{\prime})\) in the black hole formation phase indicates maximal entanglement between the two qubits, while its gradual suppression and eventual approach toward zero in the evaporation phase suggest that the \(d^{\prime}\)-qubit becomes effectively decoupled from the environment during evaporation.\\

\par\noindent\textbf{\textit{Bipartite Mutual Information:}}~In order to further probe the patterns of entanglement between various parts, we have also studied the mutual information between various pairs of qubits. The bipartite mutual information between two subsystems $A$ and $B$ is defined as $I_2(A:B) = S_{EE}(A) + S_{EE}(B) - S_{EE}(A \cup B)$, where $S_{EE}$ denotes the entanglement entropy.
As observed in Fig.~\ref{fig:I2dprimedI2d0} (Left) for the mutual information $I_2(d^{\prime}:d)$ between the qubits $d'$ and $d$, a strong $d'-d$ coupling $\tilde{J}$ promotes the formation of quantum correlations between these two qubits in the weak-coupling (formation) phase, whereas an increasing $d-0$ coupling $J_K$ tends to suppress this connection in the strong-coupling (evaporation) phase. In contrast, Fig.~\ref{fig:I2dprimedI2d0} (Right) for the mutual information $I_2(d:0)$ between the qubits $d$ and $0$ reveals that increasing the $d-0$ coupling $J_K$ from zero (in the weak-coupling (formation) phase) enhances the entanglement between them in the strong-coupling (evaporation) phase.

\par
As expected from the URG flow, the mutual information $I_{2}(d':0)$ between $d^{\prime}$ and $0$ qubits (Fig.~\ref{fig:I2dprime0I2dprime1} (Left)) $I_{2}(d':0)$ is zero in the weak-coupling phase, and develops a sharp peak precisely at the transition, subsequently decaying with increasing $J_K/\tilde{J}$ in the strong-coupling. Interestingly, this non-monotonic behavior indicates the emergence of non-local correlations between the black hole interior and near-boundary environment in the evaporation phase. A similar qualitative trend is observed for correlations between the $d^{\prime}$-qubit and qubits on other bath sites: we present $I_{2}(d':1)$ in Fig.~\ref{fig:I2dprime0I2dprime1} (Right). Note, however, that the magnitude of $I_2(d^{\prime}:i),~(i\geq 1)$ decreases with increasing $i$, reflecting the spatial decay of quantum information shared between the black hole interior and the bath with distance.
\par
These bipartite mutual informations can be viewed as an analogue of the entropy vs.\ time curves expected during black hole evaporation (see Fig.~\ref{fig:HawPageCurve} and the related discussion in Sec.\ref{introsec}). Recall that the ratio $J_K/\tilde{J}$ can be interpreted as a tuning parameter that controls the degree of coupling of the quantum system to the environment vis-a-vis within itself (i.e., between the $d$ and $d'$ qubits).

The mutual information $I_2(d':d)$ decreases within the strong-coupling phase, analogous to the weakening of shared quantum information between the black hole degrees of freedom during evaporation. 
On the other hand, $I_2(d:0)$ increases within the strong coupling phase, analogous to the growth of quantum information shared between the black hole boundary degrees of freedom and their environment during evaporation (and reflected in the increase of entropy in Hawking radiation~\cite{hawking1975particle}). This behaviour is also confirmed by studying the magnetisation of the $d'$, $d$ and bath $0$th site qubits (see Appendix~\ref{appMagnetization} for a detailed discussion).\\

\par\noindent
\textbf{\textit{Logarithmic Negativity:}}~The mutual information, $I_2(A:B)$, measures the total correlations between two subsystems, $A$ and $B$. However, it provides only an upper bound on the entanglement between them, as it does not distinguish between quantum and classical correlations. We therefore consider logarithmic negativity~\cite{vidal2003g} in order to characterize the transfer of quantum entanglement alone. Consider a system in a pure state consisting of three subsystems: $A$, $B$, and $C=(A\cup B)^c$ representing the rest of the system. The reduced density matrix describing the bipartite mixed state of subsystem $AB$ is obtained by tracing out subsystem $C$: \( \rho_{AB}=\operatorname{Tr}_C(\rho_{ABC}). \) The logarithmic negativity is then defined as \( E_N(\rho_{AB}) = \log_2\left|\left|\rho_{AB}^{T_B}\right|\right|_1, \) where $(\cdot)^{T_B}$ denotes the partial transpose with respect to subsystem $B$, and $||\cdot||_1$ represents the trace norm. The partial transpose operation transposes the indices associated with subsystem $B$, while leaving those associated with subsystem $A$ unchanged: in the product basis $\{|e_i^{(A)}\rangle\otimes|e_j^{(B)}\rangle\}$, the matrix elements of the partially transposed density matrix are given by \(  \left\langle e_i^{(A)} e_j^{(B)} \middle| \rho_{AB}^{T_B} \middle| e_k^{(A)} e_l^{(B)} \right\rangle = \left\langle e_i^{(A)} e_l^{(B)} \middle| \rho_{AB} \middle| e_k^{(A)} e_j^{(B)} \right\rangle. \)  

For our system, the logarithmic negativity of $\rho_{d^\prime d}$, $E_N(\rho_{d^\prime d})$, takes the value 1 before the transition ($J_K/\tilde{J}<1$), indicating maximal quantum entanglement between the $d$ and $d^\prime$ qubits in the formation phase. As shown in Fig. \ref{fig:LogNegativity},  it decreases to zero beyond the transition, indicating the absence of detectable quantum entanglement between the two qubits in the evaporation phase. In contrast, the logarithmic negativity of $\rho_{d0}$, $E_N(\rho_{d0})$, exhibits the opposite behaviour: it is zero in the formation phase and approaches 1 in the evaporation phase, indicating an increase in quantum entanglement between the $d$ and $0$ qubits as the system moves deeper into the evaporation phase (Fig. \ref{fig:LogNegativity}). Indeed, the variation of $E_N(\rho_{d^\prime d})$ and $E_N(\rho_{d0})$ with $J_{K}/\tilde{J}$ mirrors that of $I_{2}(d^{\prime}:d)$ and $I_{2}(d:0)$ observed in Fig.\ref{fig:I2dprimedI2d0}. In this way, $E_N(\rho_{d^\prime d})$ and $E_N(\rho_{d0})$ efficiently demonstrate the transfer of quantum entanglement from the $d^\prime$-$d$ pair to the $d$-$0$ pair. Further, the non-monotonic behaviour of $I_2(d':0)$ and $I_{2}(d':1)$ (Fig.\ref{fig:I2dprime0I2dprime1}) and $E_N(\rho_{d^\prime d})$ and $E_N(\rho_{d0})$ (Fig.\ref{fig:LogNegativity}) across the transition observed in Fig.~\ref{fig:I2dprime0I2dprime1} suggests a closer investigation of multi-partite quantum information in this system is warranted. We turn, therefore, to a computation of the tripartite mutual information (TMI) next.
\par
\noindent\textbf{\textit{Tripartite Mutual Information (TMI):}}~The tripartite mutual information (TMI) among three subsystems A, B and C is defined as $I_3(A:B:C) = I_2(A:B) + I_2(A:C) - I_2(A:BC)$.
As shown in Fig.\ref{fig:Tripartite}, the TMI among the $d^{\prime}$, $d$, and the $0$th bath site qubits ($I_3(d^{\prime}:d:0)$) for our minimal model exhibits a non-monotonic behaviour. Notably, $I_3(d^{\prime}:d:0)$ is consistently negative at and after the transition ($J_{K}/\tilde{J}\geq 1$); this indicates the monogamy of entanglement~\cite{caceffo2023negative}. The monogamy implies that, in such systems, quantum correlations dominate over their classical counterparts. Importantly, negativity of the TMI can be thought of as an alternative descriptor of scrambling dynamics~\cite{hosur2016chaos}: this characterizes the spreading of quantum information, and is closely connected to the decay of out-of-time-order correlators (OTOCs). We will discuss about scrambling and OTOC in Sec.\ref{scramsection}.
\par
A consistently negative TMI is noted as a characteristic of theories obeying the holographic duality~\cite{hayden2013holographic}; Fig.\ref{fig:Tripartite} thus hints at a possible holographic interpretation of our model. Furthermore, the non-monotonic nature of $|I_3(d':d:0)|$ for $J_{K}/\tilde{J} \geq 1$ qualitatively resembles the Page curve in the evaporation phase \cite{braunstein2013better} (see Fig.\ref{fig:HawPageCurve}), with the equivalent of the ``Page time" (i.e., the maximum of $|I_{3}(d':d:0)|$) reached for $J_{K}/\tilde{J}$ very close to the transition and followed by a gradual decay for $J_{K}/\tilde{J}>>1$. Interestingly, an entanglement phase transition was similarly observed to precede the Page time in Ref.\cite{kehrein2024}. We stress that we are drawing a heuristic analogy here between the coupling ratio $J_{K}/\tilde{J}$ and time, and not a literal mapping: Figs. \ref{fig:EntEntropy}-\ref{fig:Tripartite} display evolution of various entanglement measures under renormalisation and not temporal evolution. In Sec.\ref{scramsection}, we will study temporal evolution of the same entanglement measures in the strong coupling phase, and receive insights that are complementary to those discussed above. This convergence permits us to draw a closer analogy to the physics of black hole evaporation:~the nonlocality evidenced by the TMI in Fig.\ref{fig:Tripartite} is reminiscent of the ``island" construction obtained via holographic duality~\cite{almheiri2021entropy, almheiri2020replica, penington2022replica, almheiri2019entropy}, i.e., the quantum correlations between a region in the interior of a black hole and the Hawking radiation outside the event horizon, observed as being central to a resolution of the black hole information paradox.\\


\section{Entanglement Temperature}\label{tempsection}
In order to make further contact with black hole thermodynamics, we now calculate an ``effective temperature" arising from the hybridisation of the stationary qubits with the fermionic environment. Following similar constructions that have been explored in earlier works~\cite{kumar2016quantum, levine2016entanglement, pal2015entanglement, azuma2025black}, we develop further the notion of an entanglement temperature defined in analogy with the first law of thermodynamics. In particular, Ref.~\cite{azuma2025black} derived an analogous temperature scale from the change in coherent information (a quantity closely related to mutual information), providing a precedent for the extraction of thermodynamic quantities from changes in information-theoretic quantities. We, therefore, define a relative entanglement temperature as the change in the ground state energy $\Delta E$ of the complete system in response to a change in the mutual information ($\Delta I_{2}$) of the non-local bipartite mutual information $I_2(d':0)$ (i.e., a measure of the subsystem-bath entanglement) 
\begin{equation}
T_{\mathrm{EN}} = \frac{\Delta E}{\Delta I_2}~.
\end{equation}
We note that the changes $\Delta E$ and $\Delta I_{2}$ are measured with respect to a reference point taken just before the transition, i.e., in the weak-coupling (formation) phase where the ground state energy corresponds to the bound state energy of the singlet state due to $\tilde{J}$ together with the kinetic energy of a decoupled electronic bath, and with $I_{2}(d':0)=0$. 
$T_{\mathrm{EN}}$ establishes a direct connection between thermodynamics and entanglement transfer within the setting of unitary dynamics: 
based on the non-local mutual information $I_2(d':0)$, it effectively captures 
the flow of energy and quantum information between the interior of the subsystem (the $d'$-qubit) and the environment (the $0$-qubit).

\begin{figure}
	\centering
    \includegraphics[scale=0.32]{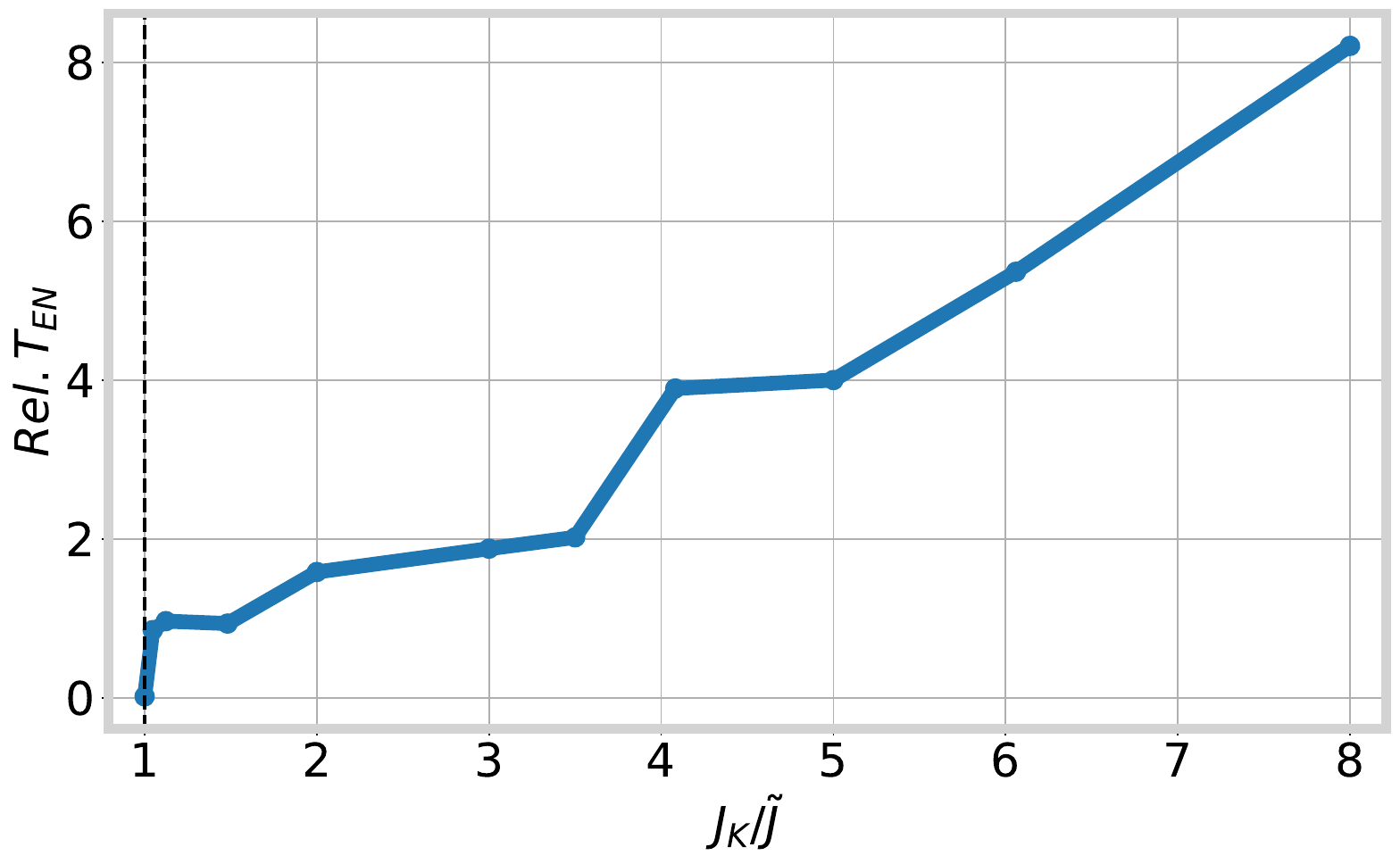}
    \caption{\raggedright Variation of relative entanglement temperature ($T_{\mathrm{EN}}$) with varying coupling ratio $J_{K}/\tilde{J}$. See main text for discussion.}
    \label{fig:EntTemp}
\end{figure}

As shown in Fig.~\ref{fig:EntTemp}, the relative entanglement temperature $T_{\mathrm{EN}}$ increases linearly as the tuning parameter $J_K/\tilde{J}$ is varied from the quantum phase transition into the strong-coupling (evaporation) phase. This reveals the importance of a temperature-like notion emergent from the transfer of quantum information between subsystem and bath due to a dissipative coupling $J_{K}$~\cite{caldeira1983quantum,chakravarty1984dynamics,guinea1985bosonization,leggettRMP1987}. The steady growth of $T_{EN}$ with varying $J_{K}/\tilde{J}$ resembles that of
the Hawking temperature of a black hole with respect to the inverse of its mass ($1/M$):
$T_H = \frac{\hbar c^3}{8\pi G M K_B} \approx \frac{10^{23}}{M}$~,
such that $T_{H}$ increases as the black hole mass $M$ decreases during evaporation. 
This requires that the coupling $\tilde{J}$ in our minimal model be interpreted as analogous to the black hole mass $M$. As the mass of the black hole decreases with time during evaporation, the growth of $T_{H}$ with time is analogous to the increase of $T_{EN}$ with the coupling ratio $J_{K}/\tilde{J}$. We stress once again that we are drawing a heuristic analogy here, and not a literal mapping: $T_{EN}$ is emergent from evolution under renormalisation and not temporal evolution. Importantly, this finding correlates well with our demonstration in Sec.\ref{scramsection} of thermalization and scrambling of various correlators and measures of entanglement with time within our minimal model.


\section{Non-Fermi Liquid Excitations and Entanglement Spreading into the Bath}\label{nflsection}
Which low-energy excitations arising from the qubit subsystem's interaction with the fermionic environment are responsible for the entanglement transfer? Some answers for the case of black hole evaporation have already been offered in the literature. Specifically, by employing a holographic duality between models of strongly interacting quantum matter and theories of weakly coupled gravity, the emergence of non-Fermi liquid (NFL) physics in the former has been connected to the physics of black holes in the latter~\cite{liu2012black,lee2009non,cinti2026holographic} (see \cite{anderson2013strange} for a contrarian viewpoint). 
The holographic dimension represents a coarse-grained energyscale of the interacting quantum system: the duality connects the evolution under renormalisation of the quantum theory situated at the UV boundary of the dimension with the dynamics of the gravitational system in the bulk, with the deep IR limit corresponding to the horizon of the black hole. Specifically, the non-Landau quasiparticle nature of the low-energy degrees of freedom in the strongly coupled quantum matter theory on the boundary is mapped onto the 
 spacetime fluctuations observed near the event horizon. This then offers another way by which to connect the study of our minimal qubit model for interacting quantum matter to the physics of black holes. We find that the transition in the minimal model from the $d^\prime$-$d$ dominated regime to the $d$-bath dominated regime produces strongly correlated low-energy excitations involving the subsystem and the fermionic environment are collective in nature. 
 The holographic duality prescription then suggests that the NFL behaviour driving 
 the information transfer during the evaporation process
 correspond to excitations that are beyond the independent quasiparticle paradigm. We demonstrate below the NFL behaviour within our minimal model using two complementary approaches: (i) a perturbative effective Hamiltonian analysis of the URG fixed point, and (ii) a non-perturbative treatment of the spectral function of the $d$-qubit obtained from the URG flow.\\

\par\noindent{\textbf{\textit{Effective Hamiltonian for gapless excitations:}}}~We begin by considering the zero-bandwidth limit of the Hamiltonian eq.\eqref{bareHam} (see Appendix~\ref{appZero} for details), and treat perturbatively the coupling to the electronic environment. The conduction bath modeled as a one-dimensional system of non-interacting electrons hopping on a lattice,
\(
-t \sum_i \left(c_i^\dagger c_{i+1} + \text{h.c.}\right)
\)~.
Within this framework, we use the fixed-point value of the subsystem-bath coupling $J_K^*$ obtained from the URG analysis to track the quantum phase transition discussed above in Sec.\ref{urgsection}. On the weak-coupling side of the transition, $J_K^* = 0$, leading to a four-fold degenerate ground state in which the qubits $d'$ and $d$ form a singlet, while the $0$th site of the conduction bath hosts either a holon, a spin-up electron, a spin-down electron or a doublon. At and beyond the transition on the strong-coupling side, the presence of a finite $J_K^*$ leads to a situation analogous to the zero-bandwidth Hamiltonian with non-zero $J_K$, exhibiting a two-fold degenerate ground state (see Appendix~\ref{appZero}). Similarly, for the special case of $\tilde{J} = 0$ (i.e., the extreme strong-coupling limit), the system again shows a two-fold degeneracy, where the $d$-qubit forms a singlet with the qubit on the $0$th bath site, and the decoupled qubit $d'$ corresponds to a doubly degenerate free spin-$1/2$ moment.
\par
Our goal is to understand how these degenerate ground states of the fixed-point zero-mode Hamiltonian evolve when the $d'-d-0$ qubit subsystem is coupled to the first site of the conduction bath through electron hopping ($V = -t(c^{\dagger}_{0}c_{1} + \textrm{h.c.})$). 
In particular, we aim to determine whether the interaction with conduction electrons preserves the original ground state or induces a transition to another ground state within the degenerate manifold. The outcome of this analysis provides insight into whether the low-lying excitations of the combined system behave similar to the Landau quasiparticles of a local Fermi liquid, or whether an orthogonality catastrophe mechanism leads to the excitations of a non-local Fermi liquid.
\par
In the weak-coupling regime, the $d$-qubit and the $0$-th bath site are effectively decoupled as $J_K^* = 0$: the perturbation due to electron hopping from the $1$st bath site does not introduce any new physics here. On the other hand, in the extreme strong-coupling limit of $\tilde{J} = 0$ (corresponding to a fully Kondo-screened system for the $d$-spin moment), previous studies~ \cite{mukherjee2022unveiling,nozieres1980kondo} indicate that the system exhibits local Fermi liquid behaviour at the first site of the conduction bath~\cite{nozieres1974fermi}. Therefore, our analysis will primarily focus on the physics of the transition and strong-coupling regime obtained for $J_{K}/\tilde{J}\geq 1$.
\par
At first order, the perturbation involving electron hopping ($V$) given earlier does not contribute, as
\(
\bra{\psi} V \ket{\psi} = 0~.
\)
Hence, we consider the second-order perturbation term $V G V$, where the Green's function $G$ is defined as
\(
G = \frac{1}{E_i - E_n}~,
\)
with $E_i$ and $E_n$ representing the energies of the initial and intermediate states, respectively. We can simplify the effective Hamiltonian by introducing the notation 
$\ket{\psi_{gs1}} \equiv \ket{\Uparrow}$ and $\ket{\psi_{gs2}} \equiv \ket{\Downarrow}$, 
which span the two-fold degenerate ground-state manifold given in Appendix~\ref{appZero}. The effective Hamiltonian is then obtained as
\begin{align}
	\mathcal{H} =  \, P_{n_{d^\prime,d,0}=1} \Big[ 
	\mathcal{J}^\perp \Big(S_{gs}^- S_1^+ + S_{gs}^+ S_1^- \Big) 
	+ \mathcal{J}^z S_{gs}^z S_1^z \nonumber \\ 
	  + \mathcal{H}_0^1 S_{gs}^z 
	+ \mathcal{H}_0^2 S_1^z 
	+ C \Big]~,
    \label{effHam}
\end{align}
where the following pseudospin operators are defined in the ground-state subspace as 
$S_{gs}^+ = \ket{\Uparrow}\bra{\Downarrow}$~,~
$S_{gs}^- = \ket{\Downarrow}\bra{\Uparrow}$~,~
\(
\ket{\Uparrow}\bra{\Uparrow} = P_{n_{d',d,0}=1} \left(\frac{1}{2} + S_{gs}^z\right)~,~
\ket{\Downarrow}\bra{\Downarrow} = P_{n_{d',d,0}=1} \left(\frac{1}{2} - S_{gs}^z\right)~.
\)
The explicit expressions for the coefficients $\mathcal{J}^\perp, \mathcal{J}^z, \mathcal{H}_0^1, \mathcal{H}_0^2$ and $C$ are provided in Appendix~\ref{appCoeffHam}.
The resulting effective Hamiltonian eq.\eqref{effHam} corresponds to an anisotropic Heisenberg model that describes how the entangled ground state formed by the qubit sites $d'$, $d$, and the $0$th bath site extends its entanglement to the first site of the conduction bath. Notably, the presence of the spin-flip terms highlight scattering processes that go beyond the conventional local Fermi liquid picture: instead, they indicate the emergence of non-Fermi liquid (NFL) behaviour~\cite{si1993metallic, varma2002singular,kotliar1993quantum}. Finally, in the context of the analogy between the physics of our minimal model and the evaporation of a black hole, we note that it has been argued in the literature by using holographic duality based methods that non-Fermi liquid behaviour originates similarly from physics associated with the black hole event horizon~\cite{phillips2022stranger,sachdev2015bekenstein,faulkner2010black}. This reinforces our suspicion that a holographic interpretation exists for our minimal model.\\

\par\noindent{\textbf{\textit{Local spectral function for the $d$-qubit:}}}~We study the spectral function \(A(\omega)\) associated with the $d$-qubit by summing over the complete spectral weight from the ultraviolet (UV) to the infrared (IR) scales received through the non-perturbative URG flow. The spectral function is defined as
\begin{align}
A(\omega) = \frac{1}{d_0} \sum_{n,0} \Big[ |\langle 0|\mathcal{O}_\sigma|n\rangle |^2 \delta (\omega + E_0 - E_n) \nonumber \\
+  |\langle n|\mathcal{O}_\sigma|0\rangle |^2 \delta (\omega - E_0 + E_n)\Big]~,
\end{align}
where $\mathcal{O}_\sigma = S_d^{-\sigma} c_{0,-\sigma}$ represents a low-energy excitation operator acting on the $d$-qubit.
\begin{figure}
	\centering
    \includegraphics[scale=0.35]{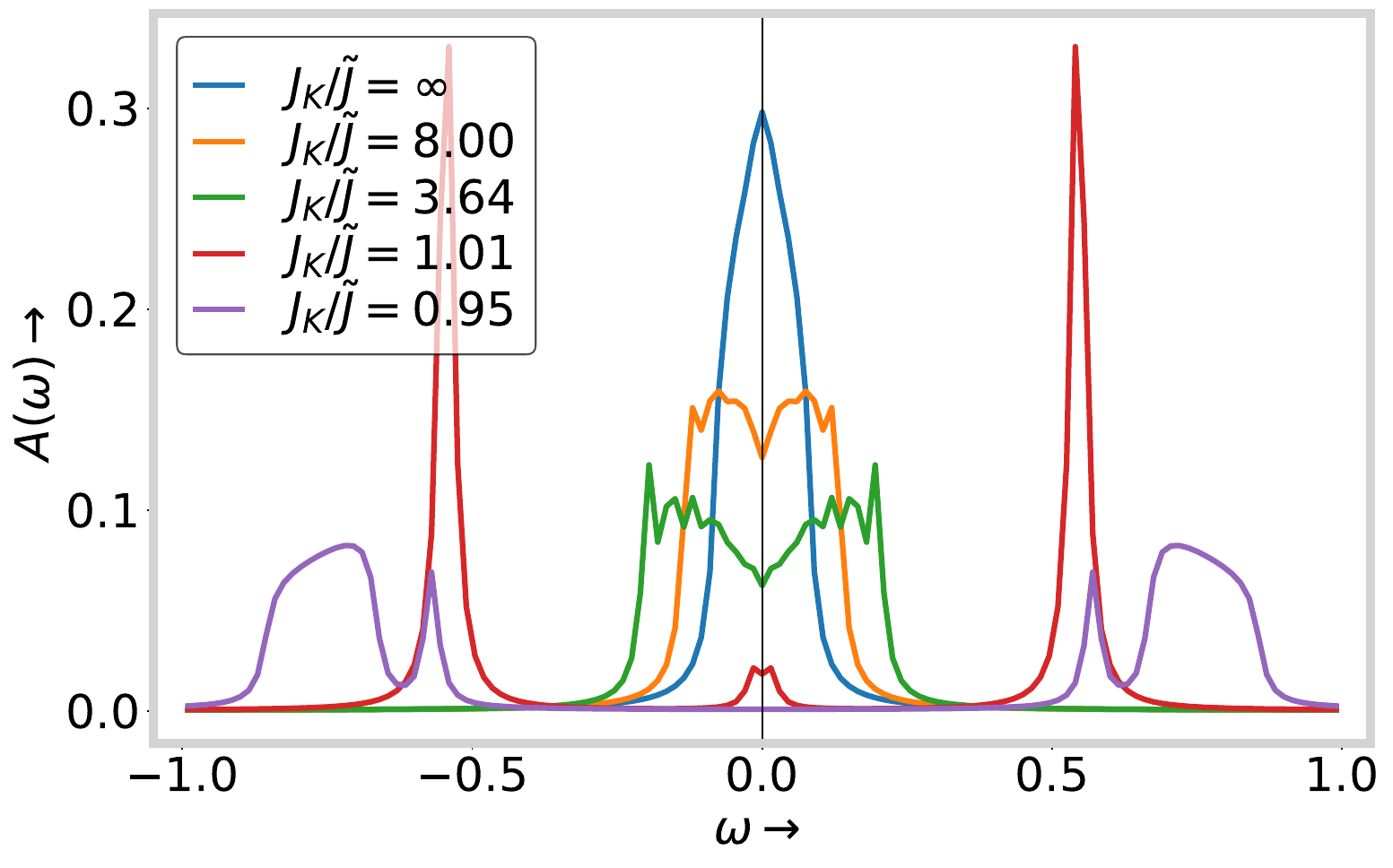}
    \caption{\raggedright Spectral function $A(\omega)$ of the $d$-qubit as a function of probe frequency $\omega$ for various coupling ratio $J_{K}/\tilde{J}$. See main text for discussion.}
    \label{fig:SpecFunc}
\end{figure}
As mentioned earlier, in the extreme strong-coupling limit $\tilde{J} = 0$, the system reduces to the conventional Kondo model. In this case, due to Kondo screening, the spectral function exhibits a sharp Kondo resonance peak at $\omega = 0$. This behaviour is recovered in our results for $J_K/\tilde{J}=\infty$ (blue curve in Fig.~\ref{fig:SpecFunc}). For a finite $\tilde{J}$ (but $J_{K}/\tilde{J}>1$), however, a suppression develops at $\omega = 0$, leading to a dip in the spectral function that signals a pseudogap-like behaviour. This pseudogapped feature persists all through the strong-coupling phase right upto the transition at $J_{K}/\tilde{J}=1$, with the depth of the dip gradually decreasing as the system approaches the critical point. In contrast, in the $\tilde{J}$-dominated weak-coupling phase ($J_K/\tilde{J} < 1$), the spectral function exhibits a clear gap at and near $\omega = 0$, indicating the absence of low-energy excitations in this regime. These spectral features provide a dynamical complement to the entanglement-based description of the transition. The suppression of low-energy spectral weight reflects
a change in the excitation spectrum of the $d$-qubit as its correlations with the bath and the $d'$-qubit
change. Although the spectral function is not itself an information-theoretic measure, its evolution helps characterize the physical processes accompanying the redistribution of quantum correlations.

\par
We have also investigated the single-particle  self-energy, defined as
$\Sigma(\omega) = G_0(\omega)^{-1} - G(\omega)^{-1}$~,
where $G_0(\omega)$ is the non-interacting Green's function and $G(\omega)$ is the interacting Green's function. In our analysis, we take $G_0$ to be the Green's function corresponding to the $\tilde{J} = 0$ limit, while $G$ corresponds to the interacting system with finite $\tilde{J}$. This allows us to isolate and study the effect of the $d'$-qubit through the self-energy. In the \(J_K\)-dominated strong-coupling phase with a pseudogapped spectral function, the low-energy behaviour of the imaginary part of the \(\Sigma^{''}(\omega)\) possesses the form
$-\Sigma^{''}(\omega)^{-1} \approx -\Sigma^{''}(0)^{-1} + \omega^2$~, matching the non-Fermi liquid nature of the excitations observed recently for a pseudogapped Mott metal~\cite{Mukherjee_2026}. 

The above results for the 
non-Fermi liquid excitations complement the information-theoretic results shown earlier for the transfer of entanglement into the bath: the entanglement and mutual-information measures track how correlations are redistributed among the qubits and the bath, while the effective Hamiltonian and spectral function characterize the accompanying low-energy dynamics. Together, these results provide a systematic setting by which to explore questions about the encoding of information and subsystem-environment correlations that arise in discussions of the black-hole information paradox.


\section{Thermalisation and Scrambling of Subsystem Information}\label{scramsection}
We now study the thermalisation of the dynamics of the local qubit degrees of freedom arising from their interaction with the fermionic bath. 

\subsection{Out-of-Time-Order Correlator (OTOC)}
The out-of-time-order correlator (OTOC) is defined as
\begin{equation}
	C(t) = - \langle [W(t), V(0)]^2 \rangle,
\end{equation}
where $V$ and $W$ are generic local Hermitian operators, and $W(t)$ is the time-evolved operator given by
\begin{equation}
	W(t) = e^{iHt/\hbar} \, W(0) \, e^{-iHt/\hbar}.
\end{equation}
For our analysis, we choose the operators corresponding to the magnetisation of the $d'$- and $d$-qubits: \(W(t) \equiv S_{d'}^{z}(t)\) and \(V(0) \equiv S_{d}^{z}\). The time evolution of  the magnetization of the qubit \(d'\) (\(S_{d'}^{z}(t)\)) can be regarded as an effective order parameter that tracks the quantum phase transition of the minimal model Hamiltonian eq.\eqref{bareHam} (see Appendix~\ref{appMagnetization} for a detailed discussion). Furthermore, since the \(d'\)- and $d$-qubits plays roles analogous to degrees of freedom of the black hole in the interior and on the boundary respectively, their time evolution likely offers insight into the dynamics of black hole evolution.

\begin{figure}
	\centering
    \includegraphics[scale=0.28]{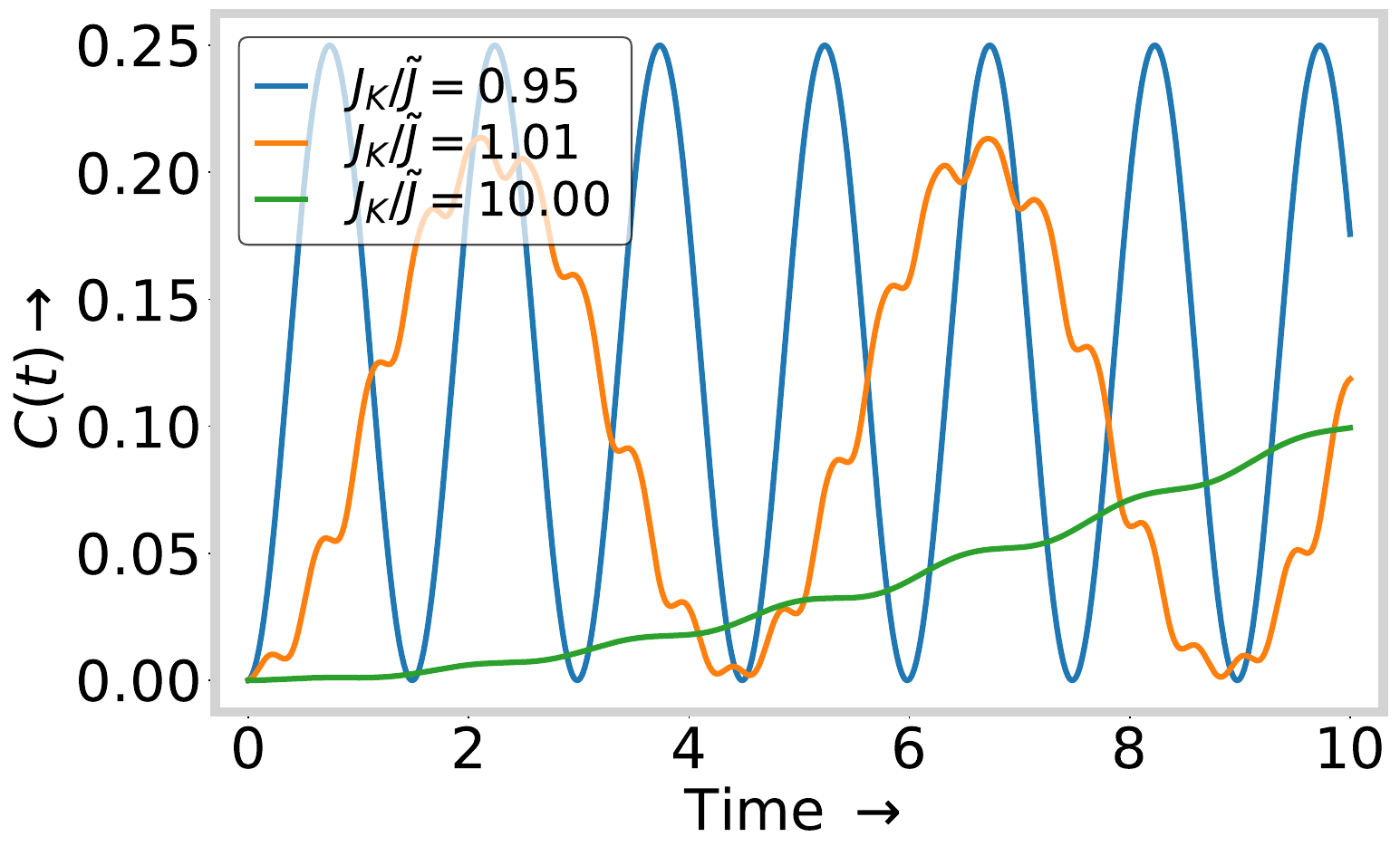}
    \includegraphics[scale=0.28]{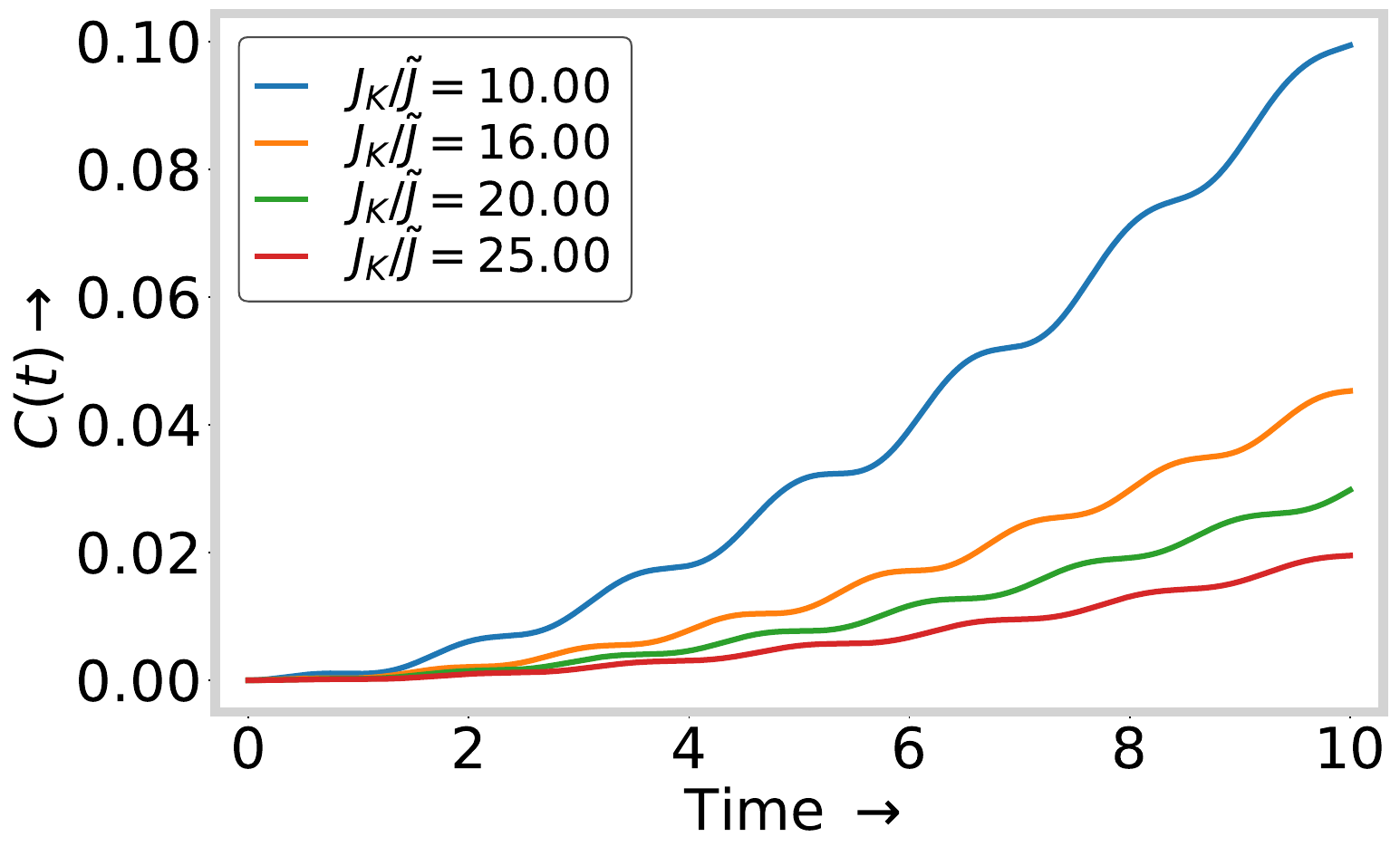}
    \caption{\raggedright (Left) OTOC for some values of the coupling ratio $J_{K}/\tilde{J}$ in the weak-coupling phase, just after the transition, and in the strong-coupling phase beyond the transition. (Right) OTOC for some values of the coupling ratio $J_{K}/\tilde{J}$ deep in the strong-coupling regime. See main text for discussion.}
    \label{fig:OTOC}
\end{figure}

Figure~\ref{fig:OTOC} (Left) shows that for $J_K/\tilde{J} = 0.95$ (weak-coupling phase), the OTOC exhibits smooth oscillatory behaviour. Just beyond the transition at $J_K/\tilde{J} = 1.01$, the oscillations become noisy: the oscillation period increases, and the height of the oscillation decreases, with time. Deep within the strong-coupling phase, the oscillation period increases significantly, indicating slower dynamical behaviour and hinting at the scrambling behaviour. Further, one can extract the Lyapunov exponent from the OTOC by fitting the early-time exponential growth~\cite{rozenbaum2017lyapunov, maldacena2016bound} to the form
\begin{equation}
	C(t) \sim -b + b\, e^{\lambda_{\mathrm{fit}} t}~.
\end{equation}
In Fig.~\ref{fig:OTOC}, we find that far inside the strong-coupling regime, the fitted Lyapunov exponent \(\lambda_{\mathrm{fit}}\) increases from \(0.19\) to \(0.22\) as \(J_K/\tilde{J}\) is varied from \(10\) to \(25\),
indicating an 
enhanced scrambling of quantum information. Given that rapid scrambling is widely regarded as a characteristic feature of black holes, this behaviour further strengthens the analogy between our minimal model and the entanglement transfer physics of black holes. We note that similar connections between scrambling dynamics and black hole physics have also been explored in earlier studies~\cite{zhang2024quantum,zhang2022quantum,liu2026scrambling, louw2024thermodynamics, pikulin2017black, kobrin2021many}.
\subsection{Time Evolution of Various Entanglement Measures}

\begin{figure}[!ht]
\centering
\includegraphics[scale=0.28]{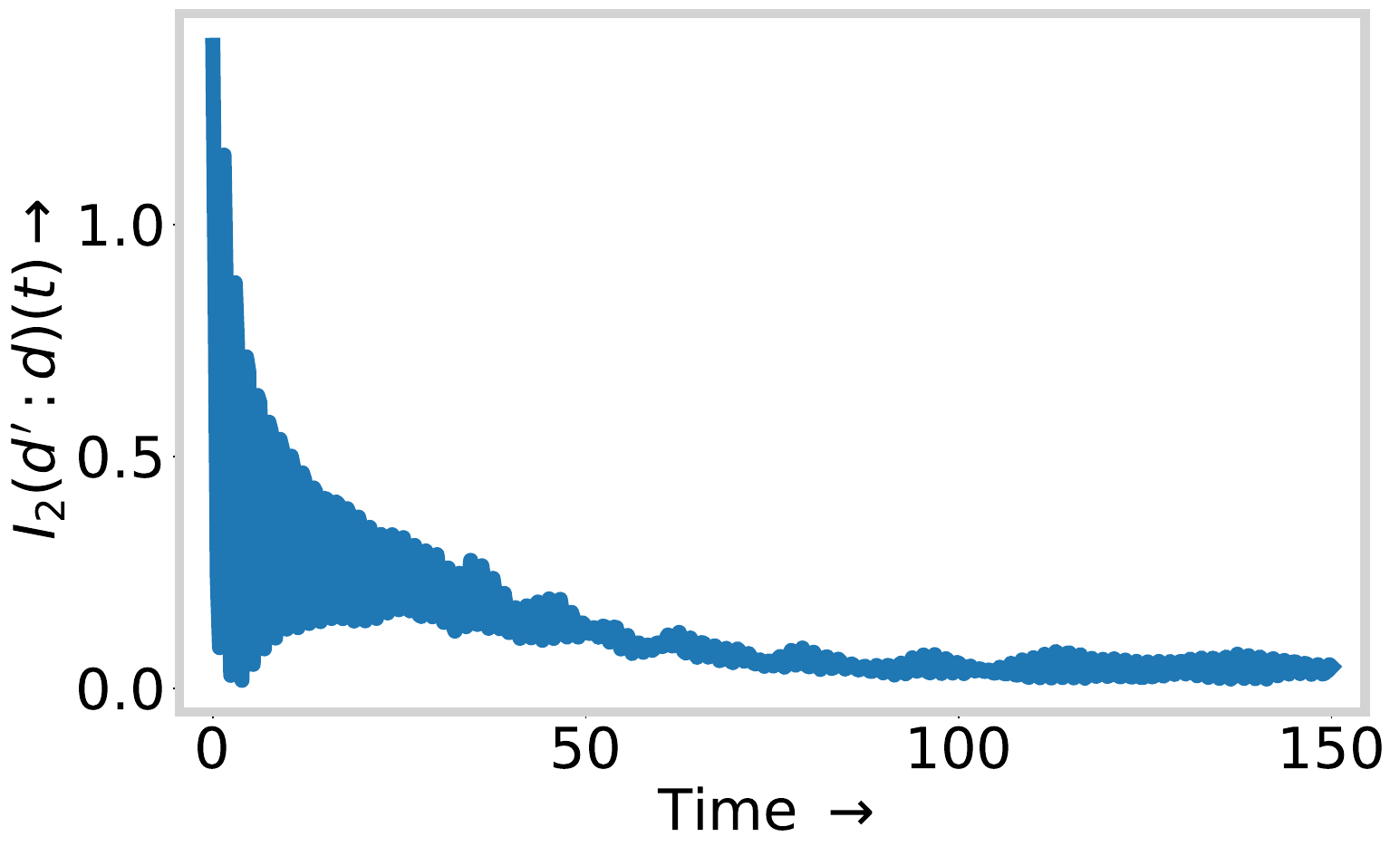}
\includegraphics[scale=0.28]{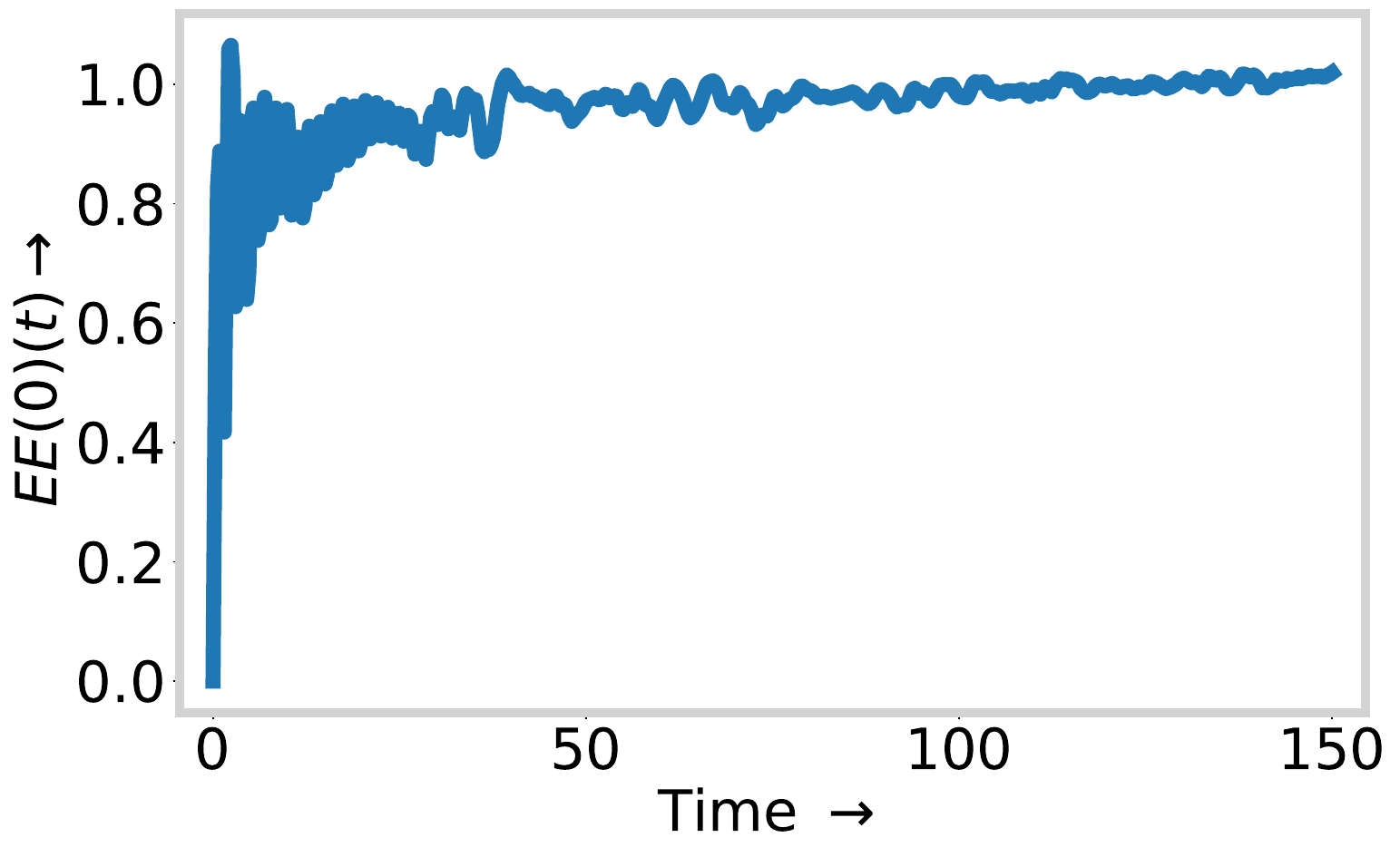}
\includegraphics[scale=0.28]{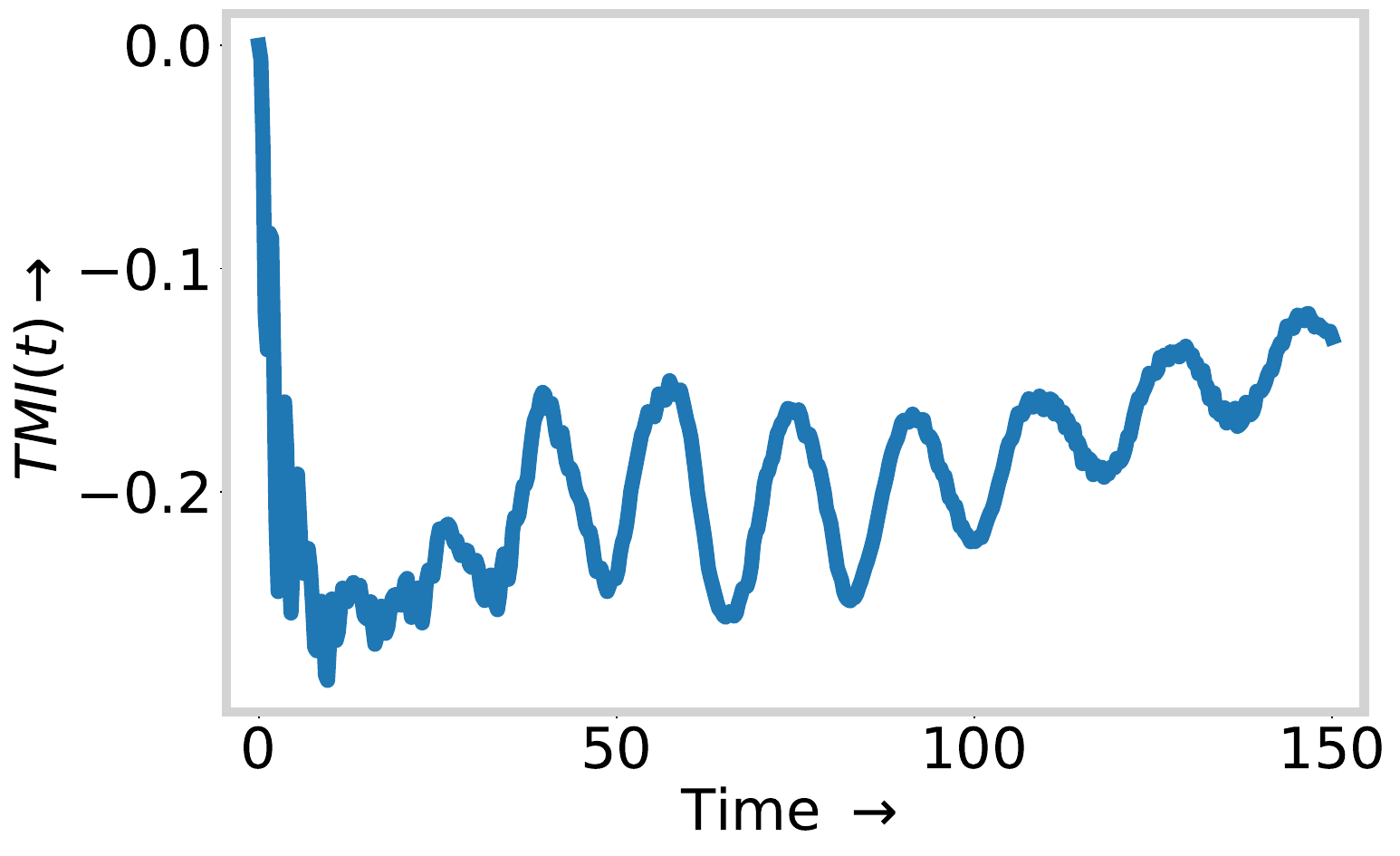}
\caption{\raggedright Time evolution of various entanglement measures: (Upper Left) mutual information between qubits $d'$ and $d$, $I_2(d':d)(t)$; (Upper Right) entanglement entropy of the bath at the $0$th site, EE$(0)(t)$; (Lower) tripartite mutual information among $d'$, $d$, and the $0$th site, TMI$(t)$. See main text for discussion.}
\label{fig:TimeEvolveEnt}
\end{figure}

We study the time evolution of various entanglement measures deep inside the strong-coupling (evaporation) phase at $J_K/\tilde{J}=4$, assuming the weak-coupling (formation) state as the initial ground state. Fig.~\ref{fig:TimeEvolveEnt} shows the time evolution dynamics of different entanglement quantities. The bipartite mutual information between the qubits $d'$ and $d$, $I_2(d':d)(t)$ (Fig.~\ref{fig:TimeEvolveEnt}, Upper Left), exhibits a decay that is qualitatively similar to the black hole entropy (Fig.~\ref{fig:HawPageCurve}). Concomitantly, the entanglement entropy of the bath at the $0$th site, EE$(0)(t)$ (Fig.~\ref{fig:TimeEvolveEnt}, Upper Right), grows with time in a manner analogous to Hawking’s calculation (Fig.~\ref{fig:HawPageCurve}). Interestingly, similar analogies were also drawn by us from Fig.~\ref{fig:I2dprimedI2d0} for the variation of $I_2(d':d)$ and $I_2(d:0)$ respectively with $J_{K}/\tilde{J}$. 
\par
Finally, the time evolution of the envelope of the modulus value of the tripartite mutual information among $d'$, $d$, and the $0$th site, TMI$(t)$ (Fig.~\ref{fig:TimeEvolveEnt}, Lower), displays a non-monotonic Page curve–like structure similar to that shown in Fig.~\ref{fig:Tripartite}, with the equivalent of the ``Page time" (i.e., the maximum of $|I_{3}(d':d:0)(t)|$) reached at a short timescale and followed by a gradual decay for longer timescales. 

All three quantities shown in Fig.\ref{fig:TimeEvolveEnt} also exhibit clear signatures of thermalization under time evolution, eventually approaching saturated values at long times. This behaviour further confirms the scrambling of quantum information between the $d'-d$ qubit subsystem and the fermionic environment during the entanglement transfer process in the strong-coupling (evaporation) phase.

\section{Conclusion}\label{concsection}

In this work, we studied a minimal model of a highly entangled quantum subsystem of two localized qubit degrees of freedom that are interacting with one another, and one of whom is coupled locally to a macroscopic fermionic environment. 
Within a unitary framework, we demonstrated how a competition between the interaction within the subsystem and that with the environment leads to the transfer of entanglement from the quantum subsystem to the environment. Our results show that several important features of the entanglement transfer bear analogies to the physics of a black hole coupled to an environment. In particular, the minimal model exhibits two distinct weak- and strong-coupling regimes that can be interpreted as analogues of the black hole formation and evaporation phases respectively; the transition between them is analogous to a continuous Hawking Page transition.
\par
We showed that suitably chosen bipartite and tripartite mutual informations as well as logarithmic negativity measures, studied both as a function of the coupling ratio $J_{K}/\tilde{J}$ (i.e., the coupling to the environment with the coupling between the two qubits of the subsystem) and as a function of time, exhibit behaviour qualitatively similar to the evolution of entropy during black hole evaporation. Additionally, a consistently negative tripartite mutual information suggests multipartite entanglement structures analogous to holographic systems and provides evidence of quantum information scrambling. We also investigated the behaviour of an effective entanglement temperature that displays similarities with the Hawking temperature associated with black hole evaporation. Furthermore, the system displays non-Fermi liquid behaviour at the quantum critical point and everywhere within the strong-coupling (evaporation) phase, drawing parallels with the physics expected near the event horizon.  Signatures of thermalisation are also observed in time evolution of a certain OTOC, whose growth in the strong-coupling phase shows an exponential growth with a Lyapunov exponent that increases slightly with increasing $J_{K}/\tilde{J}$.
\par
In summary, our results provide a complementary perspective on entanglement transfer in an open quantum system, the emergence of thermodynamics from an information perspective and its connection to physics of black hole evaporation and continuous Hawking-Page transitions. The minimal model thus serves as an efficient many-body platform for exploring information transfer, scrambling, and emergent thermodynamic behaviour within a fully unitary quantum framework, offering thereby new insights into a quantum information-theoretic resolution of the black hole information paradox. Several interesting directions for future investigations can now be pursued, and we mention only a few here. First, while we have presently computed the quantum-information based temperature scale $T_{EN}$ from the fixed point theory of the URG flow, it would be interesting to track the emergence of $T_{EN}$ directly from the URG analysis of the subsystem-bath coupling flow to strong-coupling. Such an investigation could likely yield insights on the organisational principles that guide how thermodynamics is emergent from a coarse-graining scheme that treats the fluctuations of a dissipative bath to which the quantum subsystem is coupled. Further, the thermalisation of the subsystem eigenstate with temporal evolution observed by us offers a test of the proposed equivalence between the entanglement and thermal entropies of a subsystem~\cite{abanin2019}. Finally, it will be interesting to search for a holographic realisation of a gravitational equivalent of the minimal model we have analysed here, and investigate whether its analysis reveals geometric insights into the process of entanglement transfer. A first step in this direction involves understanding the nature of the boundary conformal field theory that describes the QPT we observed above, and the emergence of the non-Fermi liquid excitations presented here. This will be reported elsewhere.


\section*{Acknowledgements}
S.L. thanks the SERB, Govt. of India for funding through MATRICS grant MTR/2021/000141 and Core Research Grant CRG/2021/000852. S.L. also thanks the Anusandhan National Research Foundation, Govt. of India for funding through Advanced Research Grant ANRF/ARG/2025/004414/PS. D.D. and A.M. thank IISER Kolkata for funding through JRF and SRF positions. The authors gratefully acknowledge discussions with R. K. Nayak, R. K. Singh, A. Dasgupta, D. Mondal and A. Das. The authors are thankful to an anonymous referee for comments.
\par
\begin{center}
\textbf{AI Usage Statement}
\end{center}
\par
All authors state that AI has been used only for improvement of language, and checking of grammar. No AI has been used for statement of the problem, analysis etc.

\begin{appendices}

\section{URG Analysis of the Minimal Model}
\label{appURGDetail}
For a detailed description of the URG formalism, we refer the reader to the Methods subsection of Ref.~\cite{debata2026kondo}. Here, we present only the details of the URG calculation for the minimal model Hamiltonian eq.\eqref{bareHam}. 

The diagonal part of the Hamiltonian can be written as $\sum_q \Big( \varepsilon_q \tau_{q\sigma} + \frac{J_K^{z}}{2} S_d^z c_{q\uparrow}^\dagger c_{q\uparrow} - \frac{J_K^{z}}{2} S_d^z c_{q\downarrow}^\dagger c_{q\downarrow} + \tilde{J} S_{d^\prime}^z S_d^z \Big)$ where $\tau_{k\sigma}=n_{k\sigma} -\frac{1}{2}$. On the other hand, the off-diagonal part of eq.\eqref{bareHam} can be written in terms of scattering processes that involve particle-like excitations

\begin{align}
    H_1^I = \sum_{\abs{k}<\Lambda,q} \frac{J_K^{z}}{2} S_d^z c_{k\uparrow}^\dagger c_{q\uparrow} - \frac{J_K^{z}}{2} S_d^z c_{k\downarrow}^\dagger c_{q\downarrow} \nonumber \\
    +\frac{1}{2} \sum_{\abs{k}<\Lambda,q} J_K^{\perp}\Big(S_d^+ S_{kq}^- + S_d^- S_{kq}^+ \Big)~, 
\end{align}

and those that involve hole-like excitations
\begin{align}
    H_0^I = \sum_{\abs{k}<\Lambda,q} \frac{J_K^{z}}{2} S_d^z c_{q\uparrow}^\dagger c_{k\uparrow} - \frac{J_K^{z}}{2} S_d^z c_{q\downarrow}^\dagger c_{k\downarrow} \nonumber \\
    +\frac{1}{2} \sum_{\abs{k}<\Lambda,q} J_K^{\perp}\Big(S_d^+ S_{qk}^- + S_d^- S_{qk}^+ \Big)~. 
\end{align}
In the above, we consider scattering processes from high momentum $q$ to the Fermi surface, $\Lambda $ is the maximum possible momentum due to a bandwidth $D$. The change in Hamiltonian at the $j$-th step of the URG is given by
 \begin{align}
 \Big(\Delta H_(j) \Big)_{\vec{q},\beta} = H_{(j-1)} - H_{(j)} = c_{\vec{q},\beta}^\dagger T_{\vec{q},\beta} \frac{1}{\omega - H_D} T_{\vec{q},\beta}^\dagger c_{\vec{q},\beta} + c_{\vec{q},\beta} \frac{1}{\omega - H_D} c_{\vec{q},\beta}^\dagger T_{\vec{q},\beta}~,
 \end{align}
where $T_{\vec{q},\beta}$ and $T_{\vec{q},\beta}^{\dagger}$ contain that part of the Hamiltonian which does not commute with $n_{q\beta}$~\cite{mukherjee2020holographic-A}.
\par\noindent
Following the computations of Refs.\cite{debata2026kondo}, the contributions of excitations from both particle and hole sectors obtain the RG equation for the couplings $J_{K}^{z}$ and $J_{K}^{\perp}$ as

\begin{eqnarray}
 \frac{\Delta J_K^{\perp}}{\Delta D} &=& - \frac{1}{2} J_K^{z} J_K^{\perp}  (\frac{1}{w-E_{1}} + \frac{1}{w-E_{2}}) N(D)~,\\
 \frac{\Delta J_K^{z}}{\Delta D} &=& -\frac{(J_K^{\perp})^2}{2} (\frac{1}{w-E_{1}} + \frac{1}{w-E_{2}} ) N(D)~,
\end{eqnarray}
where $E_{1}= \frac{D}{2} -\frac{J_K^{z}}{4} +\frac{\tilde{J}}{4}$ and  $E_{2}= \frac{D}{2} -\frac{J_K^{z}}{4} -\frac{\tilde{J}}{4}$.

As the interaction is $SU(2)$ symmetric in our minimal model, $J_K^{\perp}= J_K^z = J_{K}$, such that the RG equation now obtains
\begin{align}
    \frac{\Delta J_K}{\Delta D} = -\frac{(J_K)^2}{2} (\frac{1}{w-E_{1}} + \frac{1}{w-E_{2}} ) N(D)~, 
\end{align}
as given in eq.\eqref{kondorg}.
\section{Magnetization of the $d'$, $d$ and Bath $0$th Site Qubits}
\label{appMagnetization}

\begin{figure}[!ht]
	\centering
    \includegraphics[scale=0.28]{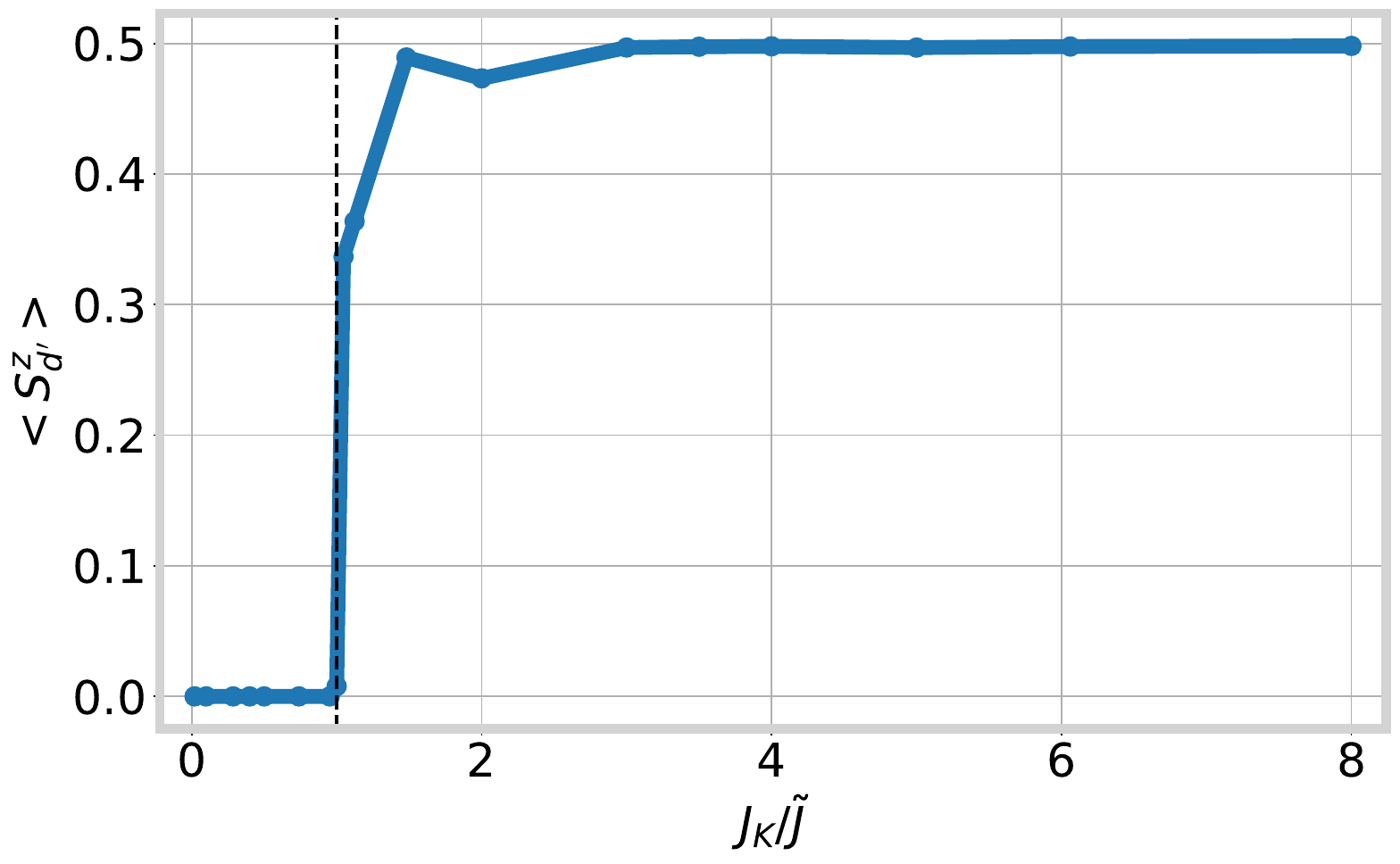}
    \includegraphics[scale=0.28]{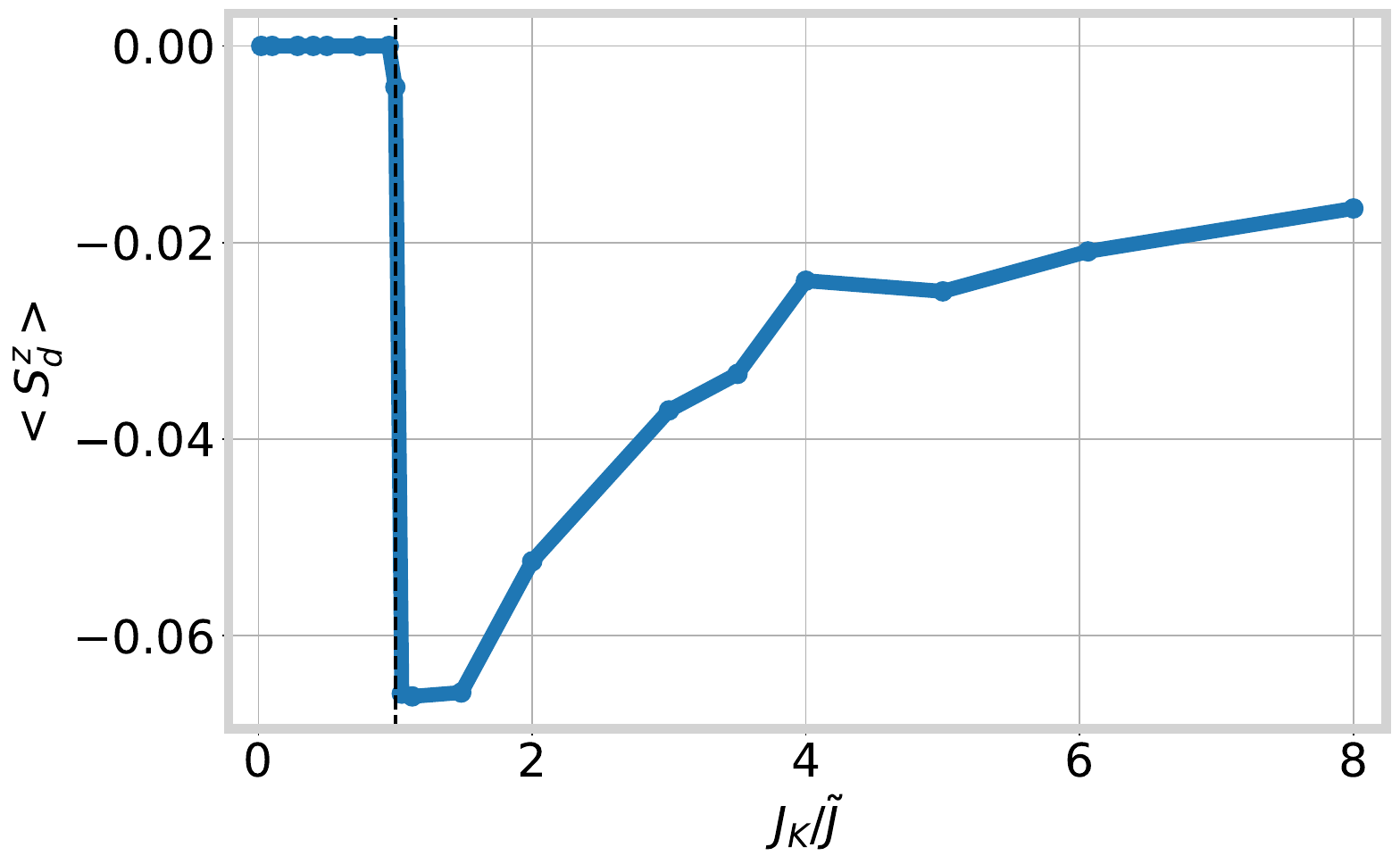}
    \caption{\raggedright (Left) Magnetisation of $d^{\prime}$-qubit and (Right) Magnetisation of $d$-qubit as a function of the coupling ratio $J_{K}/\tilde{J}$. See main text for discussion.}
    \label{fig:dprimeMagnetisationd}
\end{figure}

In the weak-coupling phase of the minimal model, the $d$ and $d^{\prime}$ qubits form a maximally entangled singlet state and the magnetisation of the $d^{\prime}$ qubit vanishes. As can be seen in Fig.~\ref{fig:dprimeMagnetisationd} (Left), an increasing $J_K$ in the strong-coupling phase progressively disrupts this entanglement; the $d^{\prime}$ qubit effectively decouples and behaves as a free spin in the extreme limit of $\tilde{J}\to 0$, i.e., it is easily polarised by a magnetic field to give the maximum magnetisation possible. Thus, the magnetisation of the $d'$-qubit, $S_{d^{\prime}}^z$, can be thought of as an order parameter of the system.
\par
In contrast, as shown in Fig.~\ref{fig:dprimeMagnetisationd} (Right), the magnetisation of the $d$-qubit remains close to zero across the entire parameter regime: it is always strongly coupled either to the $d^{\prime}$ qubit or to the $0$th site of the bath. Near the QCP, a small deviation from zero is observed, reflecting the competition between entanglement with $d^{\prime}$ and the bath’s $0$th site.
\begin{figure}
    \includegraphics[scale=0.28]{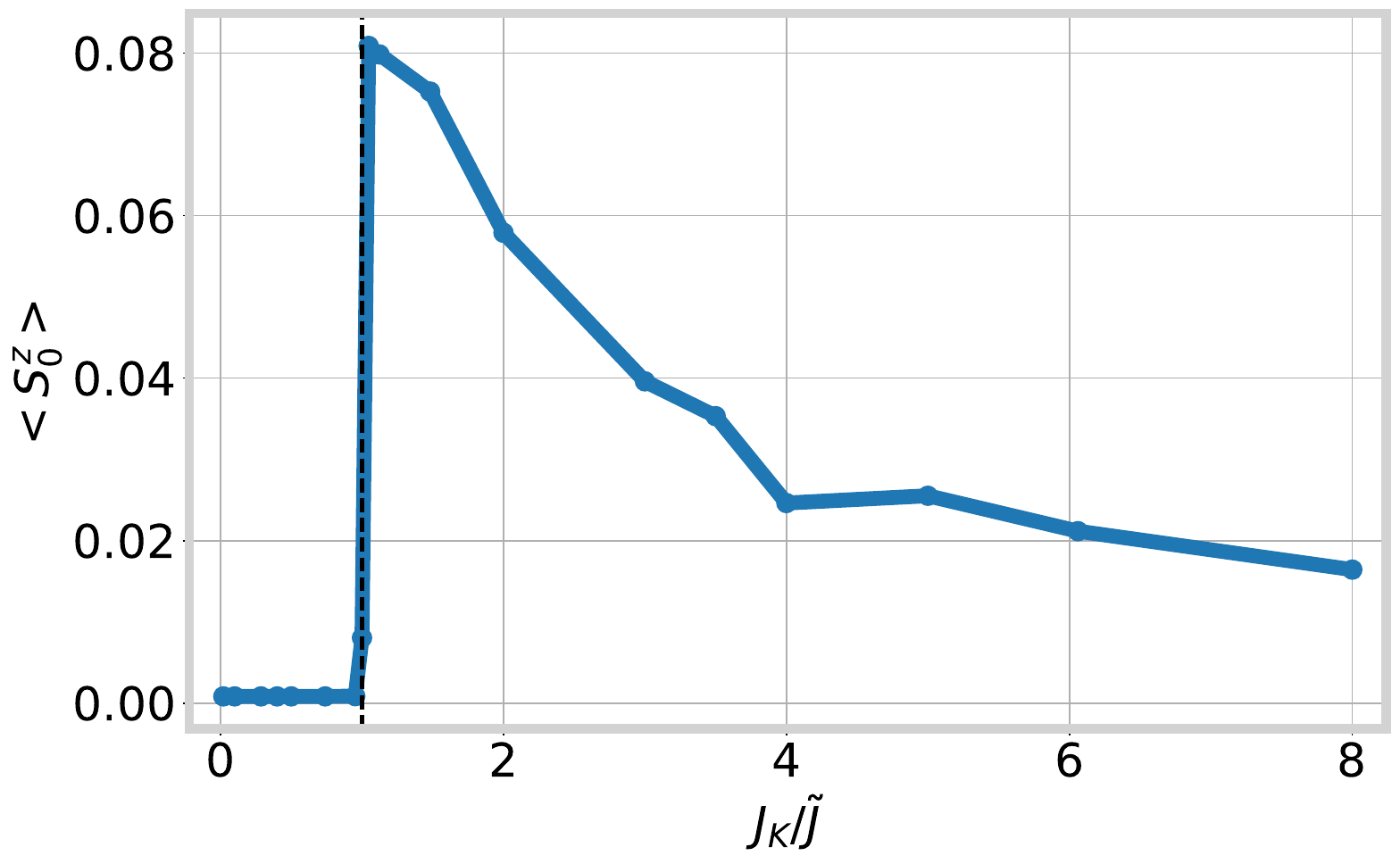}
    \centering
    \caption{\raggedright Magnetisation of the $0$th bath site as a function of the coupling ratio $J_{K}/\tilde{J}$. See main text for discussion.}
    \label{fig:0Magnetisation}
\end{figure}
\par
As seen in Fig.~\ref{fig:0Magnetisation}, the magnetisation of the $0$th bath site ($S^{z}_{0}$) is zero in the weak-coupling phase, as it is altogether decoupled from the subsystem of qubits and is only coupled through hopping to the other bath sites. In the strong coupling phase, the effective coupling $J_K^*$ dominates over $\tilde{J}$, and increases with $J_K/\tilde{J}$. Thus, for $J_K^* >> \tilde{J}$, the $d$-qubit forms a singlet with the $0$th bath site, resulting in a vanishing magnetisation at the $0$th site . There is a small jump in the height of $S^{z}_{0}$ precisely at the quantum phase transition, followed by a gradual decay within the strong-coupling phase.

\section{Zero-mode Solution}
\label{appZero}
\begin{figure}
	\centering
    \includegraphics[scale=0.35]{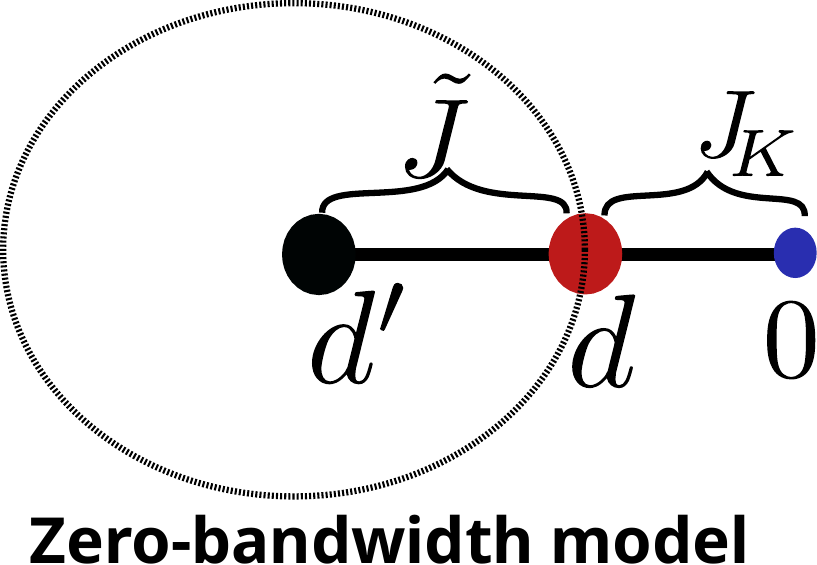}
    \caption{\raggedright Schematic diagram of the zero-bandwidth model, including only the $d'$, $d$ and bath $0$th site qubits and the couplings between them.}
    \label{fig:ZeroBandHam}
\end{figure}

The zero-bandwidth model (see Fig.~\ref{fig:ZeroBandHam}), in which the conduction bath of is effectively excluded, can be employed to gain qualitative insight into the full Hamiltonian~(see, e.g., Ref.\cite{debata2026kondo}). This simplification allows for straightforward diagonalization, enabling either an analytical or a numerically exact understanding of the system. In contrast, the full model with a finite conduction bath is significantly more complex and, in general, cannot be diagonalized exactly. 

Thus, the zero-mode Hamiltonian we consider is given by: $H  =J_K S_d \cdot S_0 + \tilde{J} S_{d^\prime} \cdot S_d $~, where $S_0$ is $S(\vec{r}=0)$. We rewrite the Hamiltonian in terms of the dimensionless coupling $g\equiv J_K^{*}/\tilde{J}$ as: 
\begin{align}
H'= \frac{H}{\tilde{J}} = g \Big[S_d^z S_{0}^z + \frac{1}{2} \Big(S_d^+ S_{0}^- +S_d^- S_{0}^+ \Big) \Big] \nonumber \\
+ [ S_d^z S_{d^\prime}^z + \frac{1}{2} \Big(S_d^+ S_{d^\prime}^- +S_d^- S_{d^\prime}^+ \Big) ]~. 
\end{align}
We will now determine the ground state eigenfunction and energy for the zero-bandwidth model. We note that the operator $S_{total}^{z}=S_{d'}^{z}+S_{d}^{z}+S_{0}^{z}$ commutes with the zero-mode Hamiltonian: the eigenvalues of $S_{total}^z$ act as a good quantum number. Thus, for the sector $S_{total}^z = \frac{3}{2}$,the lone eigenfunction $\ket{\uparrow_{d^\prime} \uparrow_d \uparrow_0}$ has the eigenvalue $(g+1)/4$. For the sector $S_{total}^z = \frac{1}{2}$, we find the eigenvalues $ \lambda_1 = (g+1)/4$, $ \lambda_2 = \frac{-2 \sqrt{g^2 -g +1} -g-1}{4}$ and $ \lambda_3 = \frac{2 \sqrt{g^2 -g +1} -g-1}{4}$, corresponding to the eigenvectors $\nu_1 = (1, 1, 1)$, $\nu_2 = (- \frac{-1 +g - \sqrt{1-g+g^2}}{g}, - \frac{1 + \sqrt{1-g+g^2}}{g}, 1)$, and $\nu_3 = (- \frac{-1 +g + \sqrt{1-g+g^2}}{g}, - \frac{1 +-\sqrt{1-g+g^2}}{g}, 1)$ respectively in the basis of
$\ket{\downarrow_{d^\prime} \uparrow_d \uparrow_0}$, $\ket{\uparrow_{d^\prime} \downarrow_d \uparrow_0}$ and $\ket{\uparrow_{d^\prime} \uparrow_d \downarrow_0}$.
\par
Similarly, for the sector $S_{total}^z = -\frac{1}{2}$, we find

the same three eigenvalues $\lambda_1$, $\lambda_2$ and $ \lambda_3$ as for the sector $S_{total}^{z}=1/2$ above, but with the eigenvectors $\nu_1 = (1, 1, 1)$, $\nu_2 = (-1 +g + \sqrt{1-g+g^2}, -g - \sqrt{1-g+g^2}, 1)$, and $\nu_3 = (-1 +g - \sqrt{1-g+g^2}, -g + \sqrt{1-g+g^2}, 1)$ respectively in the basis of $\ket{\downarrow_{d^\prime} \downarrow_d \uparrow_0}$, $\ket{\downarrow_{d^\prime} \uparrow_d \downarrow_0}$ and $\ket{\uparrow_{d^\prime} \downarrow_d \downarrow_0}$.

Finally, for the $S_{total}^z = -\frac{3}{2}$, we find the eigenvalue $(g+1)/4$ for the eigenvector $\ket{\downarrow_{d^\prime} \downarrow_d \downarrow_0}$~. 

\par\noindent
Two other states in the low-energy spectrum of the zero-mode Hamiltonian are: \\
1) $\ket{\uparrow_{d^\prime} \uparrow_d}$, or $\ket{\downarrow_{d^\prime} \downarrow_d}$, or triplet between $d^\prime$ and $d$ where bath $0$th site can take double occupancy or $0$ occupancy. This gives the eigenvalue $\lambda_{4}=1/4$, and\\
2) Singlet between $d^\prime$ and $d$ where bath $0$th site can take either double or $0$ occupancy. This gives the eigenvalue $\lambda_{5}=-3/4$.

\begin{figure}
	\centering
    \includegraphics[scale=0.32]{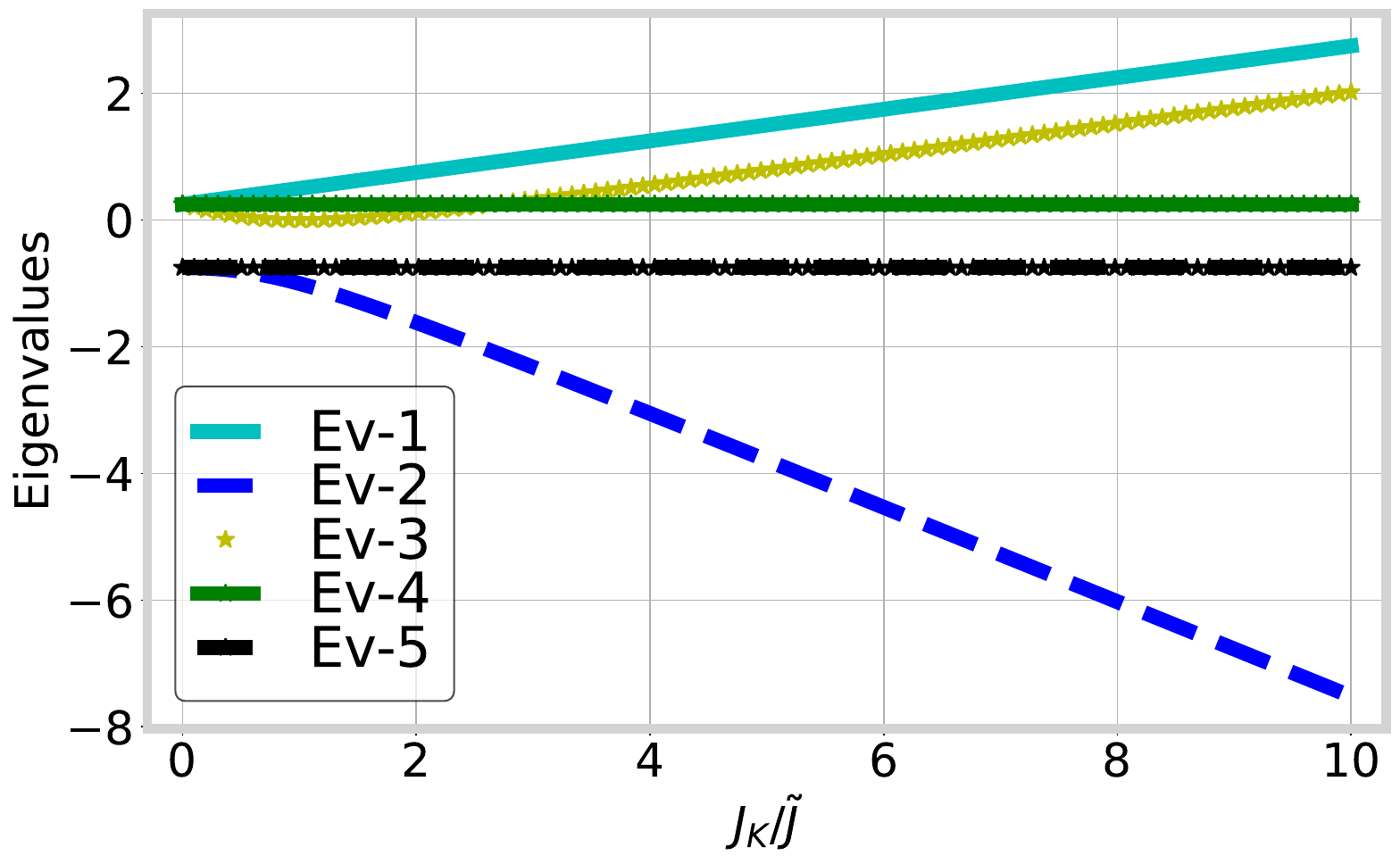}
    \caption{\raggedright Variation of eigenvalues of the zero-mode Hamiltonian with dimensionless coupling $g=J_K/\tilde{J}$. The eigenvalues Ev-1 to Ev-5 correspond to expressions for $\lambda_{1}$ to $\lambda_{5}$ given in the text.}
    \label{figZeroEigenVal}
\end{figure}
\par
We can see from Fig.~\ref{figZeroEigenVal} that $\lambda_{2}$ is the minimum eigenvalue: the results above show that the ground state has a two-fold degeneracy for $J_K/ \tilde{J} > 0$. The two degenerate ground states are from the $S_{total}^z = \pm\frac{1}{2}$ sectors:

\begin{eqnarray}
S_{total}^z = \frac{1}{2}:~&&\ket{\psi_{gs1}} = \frac{1}{\sqrt{a^2 + b^2 + c^2}}\Big[ a \ket{\downarrow_{d^{\prime}} \uparrow_d \uparrow_0} + b \ket{\uparrow_{d^{\prime}} \downarrow_d \uparrow_0} + c \ket{\uparrow_{d^{\prime}} \uparrow_d \downarrow_0} \Big]~,\\
S_{total}^z = -\frac{1}{2}:&&~\ket{\psi_{gs2}} = \frac{1}{\sqrt{j^2 + h^2 + c^2}}\Big[ j \ket{\downarrow_{d^{\prime}} \downarrow_d \uparrow_0} + h \ket{\downarrow_{d^{\prime}} \uparrow_d \downarrow_0} + c \ket{\uparrow_{d^{\prime}} \downarrow_d \downarrow_0} \Big]~,
\end{eqnarray}

where the coefficients are given by $a = - \frac{-1 +g - \sqrt{1-g+g^2}}{g}$, $b = - \frac{1 + \sqrt{1-g+g^2}}{g}$, $c = 1$, $j = -1 +g + \sqrt{1-g+g^2}$, $h = -g - \sqrt{1-g+g^2}$~.

\section{Coefficients of the Effective Hamiltonian}
\label{appCoeffHam}
The coefficients of the effective Hamiltonian for the non-Fermi liquid excitations of the strong-coupling phase, eq.\eqref{effHam}, are given by:
$\mathcal{J}^\perp = 2 \tilde{J} C_1$ and $\mathcal{J}^z=2 \tilde{J} (C_2-C_3-C_5 + C_4)$ and $\mathcal{H}_0^1 =\tilde{J}(C_2+C_3-C_4-C_5)$ and $\mathcal{H}_0^2=\tilde{J}(C_2-C_3+C_5-C_4)$ and $C=\frac{\tilde{J}(C_2+C_3+C_4 + C_5)}{2}$, where\\

$ \mathbf{C_1} = \Big(-\frac{t}{\tilde{J}} \Big)^2 \frac{ \Big( \frac{a}{\sqrt{2}}  + \frac{b}{\sqrt{2}} \Big) \Big( - \frac{h}{\sqrt{2}}  - \frac{c}{\sqrt{2}} \Big) }{\sqrt{a^2 + b^2 + c^2} \sqrt{j^2 + h^2 + c^2}} \frac{1}{E_1 - \frac{1}{4}}  +   \Big(-\frac{t}{\tilde{J}} \Big)^2 \frac{ \Big(- \frac{a}{\sqrt{2}} + \frac{b}{\sqrt{2}} \Big) \Big(\frac{h}{\sqrt{2}} - \frac{c}{\sqrt{2}} \Big) }{\sqrt{a^2 + b^2 + c^2}  \sqrt{j^2 + h^2 + c^2}} \frac{1}{E_1 + \frac{3}{4}}$~,\\

$\mathbf{C_2} = \Big(-\frac{t}{\tilde{J}} \Big)^2 \frac{c^2}{a^2 + b^2 + c^2} \frac{1}{E_1 - \frac{1}{4}} $~,~$\mathbf{C_3} = \Big(-\frac{t}{\tilde{J}} \Big)^2 \frac{\Big( \frac{a}{\sqrt{2}} +  \frac{b}{\sqrt{2}} \Big)^2 }{a^2 + b^2 + c^2} \frac{1}{E_1 - \frac{1}{4}}  +   \Big(-\frac{t}{\tilde{J}} \Big)^2 \frac{\Big( - \frac{a}{\sqrt{2}} +  \frac{b}{\sqrt{2}} \Big)^2 }{a^2 + b^2 + c^2} \frac{1}{E_1 + \frac{3}{4}}$~, \\

 $\mathbf{C_4} = \Big(-\frac{t}{\tilde{J}} \Big)^2 \frac{j^2}{j^2 + h^2 + c^2} \frac{1}{E_1 - \frac{1}{4}}$~,~$\mathbf{C_5} = \Big(-\frac{t}{\tilde{J}} \Big)^2 \frac{\Big( - \frac{h}{\sqrt{2}}  - \frac{c}{\sqrt{2}} \Big)^2 }{j^2 + h^2 + c^2} \frac{1}{E_1 - \frac{1}{4}}  +   \Big(-\frac{t}{\tilde{J}} \Big)^2 \frac{ \Big(\frac{h}{\sqrt{2}} - \frac{c}{\sqrt{2}} \Big)^2 }{j^2 + h^2 + c^2} \frac{1}{E_1 + \frac{3}{4}}$~.

\end{appendices}

\bibliography{Blackhole-Revised-1}{}
\bibliographystyle{unsrt}
\end{document}